\documentclass[]{article}

\usepackage[sort&compress,square,comma,authoryear]{natbib}

\usepackage{graphicx}
\usepackage{epstopdf, epsfig}

\usepackage{amsmath}
\usepackage{lineno}
\usepackage{siunitx}
\usepackage{subcaption}
\usepackage{ amssymb }
\usepackage{breakcites}
\usepackage[makeroom]{cancel}
\usepackage{mathrsfs}
\usepackage{upgreek}

\newcommand{\mathsfbi}[1]{\boldsymbol{\mathsf {#1}}}
\newcommand{\mathsfi}[1]{{\mathsf {#1}}}
\newcommand{\upi}{\uppi}

\newcommand{\mt}[1]{\mathrm{#1}}
\newcommand{\bs}[1]{\boldsymbol{#1}}

\newcommand{\mean}[1]{\langle #1 \rangle}

\newcommand\Tstrut{\rule{0pt}{2.6ex}}         

\newcommand{\xvec}{\boldsymbol{x}}
\newcommand{\uvec}{\boldsymbol{u}}

\newcommand{\uvecfi}{\widetilde{\boldsymbol{u}}}

\newcommand{\ufia}{\xi}

\newcommand{\uvecfia}{\boldsymbol{\xi}}

\newcommand{\qvec}{\boldsymbol{q}}
\newcommand{\qveca}{\boldsymbol{q}^*}

\newcommand{\J}{\mathscr{J}}
\renewcommand{\L}{\mathscr{L}}

\newcommand{\Real}{\operatorname{Re}}
\newcommand{\Imag}{\operatorname{Im}}

\newcommand{\NS}{\bs{\mathcal{N}}}
\newcommand{\con}{\bs{\mathcal{C}}}

\newcommand{\pfil}{p}
\newcommand{\pfila}{\pi}

\newcommand{\intt}{\int_{t_0}^{t_f}}
\newcommand{\ti}{t_0}
\newcommand{\tf}{t_f}

\newcommand{\ft}[1]{\widetilde{#1}}

\title{Reconstruction of turbulent flow fields from lidar measurements based on large-eddy simulation}
\author{Pieter Bauweraerts, Johan Meyers \\[1em]
	KU Leuven, Mechanical Engineering \\
	Celestijnenlaan 300A, B3001 Leuven, Belgium}

\begin{document}
	
	\maketitle
	
	\begin{abstract}
We investigate the reconstruction of a turbulent flow field in the atmospheric boundary layer from a time series of lidar measurements, using Large-Eddy Simulations (LES) and a 4D-Var data assimilation algorithm. This leads to an optimisation problem in which the error between measurements and simulations is minimised over an observation time horizon. To exploit the spatial coherence of the turbulence, we use a quadratic regularisation in the objective function, that is based on the two-point covariance matrix. Moreover, to improve conditioning, and remove continuity constraints, the problem is transformed into a Karhunen--Lo\`eve basis. For the optimisation, we use a quasi-Newton limited-memory BFGS algorithm combined with an adjoint approach for the gradient. We also consider reconstruction based on a Taylor's frozen turbulence (TFT) model as point of comparison. To evaluate the approach, we construct a series of virtual lidar measurements from a fine-grid LES of a pressure-driven boundary-layer. The reconstruction uses LES on a coarser mesh and smaller domain, and results are compared to the fine-grid reference. Two lidar scanning modes are considered: a classical plan-position-indicator mode, which swipes the lidar beam in a horizontal plane, and a 3D pattern that is based on a Lissajous curve. We find that normalised errors lie between $15\%$ and $25\%$ (error variance normalised by background variance) in the scanning region, and increase to $100\%$ over a distance that is comparable to the correlation length scale outside this scanning region. Moreover, LES outperforms TFT by $30\%$ to $70 \%$ depending on scanning mode and location.
	\end{abstract}
	
	\section{Introduction}
	Flow field measurements in the atmospheric boundary layer have relevance for a broad spectrum of applications, ranging from fundamental turbulent boundary layer \citep{gal1992estimations, menut1999urban}, to analysing turbulent wakes of wind turbines \citep{kasler2010wake, iungo2013, rhodes2013effect} to air quality studies \citep{wakimoto1986lidar}, among others. Pulsed lidar sensors allow taking flow field measurements almost simultaneously at multiple locations in space, at frequencies of around one Hertz, leading to a vast amount of measurement data \citep{pena2011remote}. However, lidar sensors are limited to measure a volume averaged line-of-sight wind speed.
	
	Several techniques exist for transforming this raw data into velocity vectors. A first category only deals with time-averaged flow statistics, based on the estimation of parameters of an analytical wind speed model \citep[see e.g. ][]{aitken2014quantifying, borraccino2017wind}. 
	In a second approach, the instantaneous velocity is reconstructed but only at discrete measurement points in space using simple algorithms to transform the measured wind speeds to velocities, e.g. assuming horizontal homogeneousity, and stationarity of the velocity field between the measured points in combination with doppler beam swinging lidar scan mode \citep{lundquist2015quantifying}, or assuming zero spanwise and vertical velocity in combination with a spinning lidar \citep{simley2011analysis,schlipf2013nonlinear, mikkelsen2013spinner}.
	
	All the previous approaches ignore the spatial correlations of the turbulent flow field, which are, e.g, known to extend up to $20H$ in neutrally stratified atmospheric boundary layers \citep{fang2015large}. In three-dimensional variational data assimilation (3D-Var), these correlations are taken into account \citep{lorenc1981}; a similar method named linear stochastic estimation (LSE) has been simultaneously developed in the turbulence community \citep{adrian1979conditional,adrian1988stochastic}. Recently, this has been applied to microscale flow fields \citep{krishnamurthy2013coherent,bos2016assessing,dimitrov2017application}. In these last studies, the velocity correlation is modelled using a 3D homogeneous isotropic turbulence spectral correlation tensor. This is often justified in the horizontal directions; however, in boundary layers, the vertical direction shows strong anisotropic behaviour.
	
	A method to incorporate the time series of measurements for nonlinear systems have been independently and simultaneously developed in the geoscience and control community and is respectively known as four-dimensional variatiational data assimilation (4D-Var) \citep{lewis1985use,lorenc1986analysis,le1986variational, talagrand1987variational} and nonlinear moving horizon state estimation (MHE) \citep{jang1986comparison}. This methodology minimises a linear combination of the mismatch between a time series of real and virtual observations, the model error, and the deviation from the background distribution. 
	
	The combination of large-eddy simulation (LES) and 4D-Var with lidar data to retrieve turbulent structures was first proposed in \citet{lin2001retrieval} conducting a series of twin experiments. Later in a series of papers this methodology was applied to a field measurement campaign, using two different lidars for reconstruction and validation of the methodology \citep{chai2003estimation,chai2004retrieval,newsom2004assimilating,newsom2005retrieval,xia2008retrieval}. To regularise the problem, a Laplacian-based penalty term was used. Continuity was not strictly enforced but added as an additional penalty term. The combination of 4D-Var with a Taylor's frozen turbulence (TFT) model was shown in \cite{raach2014three}, however, without including spatial regularisation. 
	
	In the current work, LES-based reconstruction of turbulence from lidar measurements is further investigated for the atmospheric boundary layer, improving on the problem formulation. To this end, regularisation of the problem is based on the background covariance tensor, following a Bayesian inference framework. Moreover, we transform the problem into a Karhunen--Lo\`eve basis (or proper orthogonal decomposition (POD) basis), which is constructed from the covariance tensor. This leads to an unconstrained optimisation problem, since the POD basis is by construction divergence free. Moreover, the problem is also optimally conditioned, removing the unfavourable scaling of the condition number of the covariance matrix with growing problem size. By using homogeneity in horizontal directions, which is generally a good approximation for atmospheric boundary layers, the POD basis can be efficiently calculated and stored, since the modes correspond to Fourier modes in horizontal directions. We test the methodology based on virtual lidar measurements in fine-grid reference simulations, using a coarser LES reconstruction grid, as well as a smaller domain. This allows us to perform detailed out-of-sample comparisons of the reconstructed turbulence with the reference solution.  
	
	The paper is further organised as follows. In \S~\ref{sec:lidar}, the lidar observations model is described. \S~\ref{sec:4D_var} discusses the state estimation methodology. Subsequently, \S~\ref{s:statespacemodels} introduces the LES and TFT models used for the state estimation. The adjoint-based optimisation methodology is discussed in \S~\ref{sec:Opt}, and details of the case setup are introduced in \S~\ref{sec:case_setup}. Results are presented in \S~\ref{sec:results}; conclusions and future outlook in \S~\ref{sec:conclusion}.

	\section{Lidar measurements} \label{sec:lidar}
	Lidar sensors measure the velocity component along the laser beam direction. Here, we focus on pulsed lidar sensors, which measure the wind speed at different locations along the beam simultaneously \citep{pena2011remote}. A much used example is the Lockheed Martin WindTracer \citep{krishnamurthy2013coherent}, which is considered in the current work as a reference. This lidar has a moderate range gate width of $\Delta r=\SI{105}{\meter}$, a pulse length (FWHM) of $\Delta p_{1/2}=\SI{105}{\meter}$, initial blind zone of $r_0=\SI{436}{\meter}$, and a total of $N_r=100$ range gates, see table~\ref{tab:LidarSetup} for a summary.
	
	Consider a lidar system mounted at location $\bs{x}_{\mt{m}}$ (extension to multiple lidar systems is straightforward, but not considered here), and a beam direction  $\bs{e}_{\mt{l}}(t)$ that follows a scanning pattern in time. The lidar measurement locations then correspond with
	\begin{equation}
	\bs{x}_{i}(t) =\bs{x}_{\mt{m}}+(r_0+\Delta r(i-1))\bs{e}_{\mt{l}}(t), ~~~~ i=1\cdots N_r.
	\end{equation}
	Due to the finite pulse, range gate width and sampling time, the space--time filtered wind speed around $\bs{x}_{i}$ is measured, oriented in the direction of the lidar beam. For a single location, we can express \citep{banakh1997estimation} 
	\begin{align} \label{eq:lidar_ms}
	h_i(\uvec(\bs{x},t),t_n) \triangleq \frac{1}{T_{\mt{s}}}\int_{t_{n-1}}^{t_n}\int_{\Omega} \uvec(\bs{x},t)\cdot \bs{e}_{l}(t) \ \mathcal{G}_{l}(\mathsfbi{Q}(t)(\bs{x}-\bs{x}_{i}(t))) \ \mathrm{d}\bs{x}\mathrm{d}t,
	\end{align}
	with $\uvec(\bs{x},t)$ the three-dimensional and time-dependent turbulent velocity field, $h_i$ the observation function, $\mathcal{G}_l$ the lidar filter kernel oriented in the direction of the lidar beam, and $\mathsfbi{Q}$ a coordinate transformation between the reference coordinate system $(\bs{e}_1,\bs{e}_2,\bs{e}_3)$, and a coordinate system aligned with the beam such that $\bs{e}_{l}=\mathsfbi{Q}\bs{e}_1$. Since the samples are only collected at discrete time instances, we introduce $t_n = \ti + nT_{\mt{s}}$, with $n$ the sample number and $1/T_s=f_s$ the sample frequency. We introduce the vector of observation functions  $\bs{h}_n=[h_{n,1},\, \ldots, h_{n,N_r}]^\top$, and similarly the vector of observations $\bs{y}_n \triangleq [y_{n,1},\,\ldots, y_{n,N_r}]^\top$ , which differ by the vector of measurement errors $\bs{w}_n$, so that $\bs{y}_n = \bs{h}_n+\bs{w}_n$. Note that the subscript $n$ is used to denote the evaluation at $t_n$, e.g. $h_{n,i}=h_i(\uvec(\bs{x},t),t_n)$.
	
	The lidar filter kernel corresponds to a convolution of a box function with width the range gate width $\Delta r$, and a Gaussian filter kernel with width $\Delta p$ \citep{banakh1997estimation,lundquist2015quantifying}. Using $\bs{x}=[x_1,\,x_2,\,x_3]^\top$, this corresponds to  
	\begin{align} \label{eq:lidar_kernel}
	\mathcal{G}_l(\bs{x}) 
	& =  \frac{1}{2\Delta p_{1/2}}\left[\mt{erf}\left(\frac{x_1 + \Delta r/2}{\Delta p_{1/e}}\right)-\mt{erf}\left( \frac{x_1 - \Delta r/2}{\Delta p_{1/e}}\right)\right]\delta(x_2)\delta(x_3),
	\end{align}
	with $\delta$ the Dirac delta function and $\Delta p_{1/e}= \Delta p_{1/2}/(2\sqrt{\log 2})$ the $1/e$ pulse width. We note that in the direction perpendicular to the beam, a Gaussian filter could be used, with a filter width that corresponds to the beam width. The maximum beam width occurs at $R_{\mt{max}}$ and can be roughly estimated by $\lambda R/d=\SI{23}{\centi \meter}$ \citep{frehlich1998coherent}, with $\lambda$ the wavelength of the lidar and $d$ the beam waist. Practical LES grid resolutions for ABL simulations typically range from $2-\SI{60}{\metre}$, such that for simplicity, the Gauss kernel in perpendicular directions is approximated by a Dirac delta function.

	\begin{table}
		\centering
		\begin{tabular}{l l l c}
			Initial blind zone\Tstrut & $r_0$ & $\SI{436}{m}$ \\
			Range gate width & $\Delta r$& $\SI{105}{m}$ \\
			Pulse length & $\Delta p_{1/2}$ & $\SI{105}{m}$ \\
			Number of range gates & $N_r$ & $100$ \\
			Maximum range & $R_{\mt{max}}$& $\SI{10831}{m}$\\
			Pulse repetition frequency & $f_{\mt{p}}$& $\SI{500}{\second^{-1}}$\\
			Sample frequency & $f_{\mt{s}}$& $\SI{5}{\second^{-1}}$\\
			Wave length & $\lambda$ & $\SI{2}{\micro\meter}$\\
			Beam waist & $d$ & $\SI{9.4}{\centi\meter}$ \\
		\end{tabular}
		\caption{Summary of the lidar parameters of a Lockheed Martin WindTracer \cite{krishnamurthy2013coherent}.}
		\label{tab:LidarSetup}
	\end{table}

	\section{State estimation approach} \label{sec:4D_var}
	\subsection{General methodology}\label{s:general_methodology}
	Given an approximate state equation, the 4D-Var algorithm provides a best estimate of the time-dependent state given a series of measurements. Here, we introduce the method, mainly following ideas from \citet{jazwinski2007stochastic, courtier1998ecmwf,lorenc1986analysis}, adopting it to the context of lidar measurements and LES-based state models and turbulent flows in the atmospheric boundary layer.
	
	Consider the (exact) space--time velocity field $\uvec(\bs{x},t)$, and an approximate state equation (e.g. the LES equations). We then have
	\begin{subequations}
		\begin{align} 
		\frac{\partial \bs{u}}{\partial t} &= \bs{f}(\bs{u},\bs{p}) + \bs{v},\label{eq:model} \\
		\uvec(\bs{x},t_0) &= \bs{u}_0(\bs{x}), \label{eq:initial}
		\end{align}
	\end{subequations}
	where $\bs{f}$ is a shorthand notation for the momentum balance in the approximate flow model,  $\bs{v}(\bs{x},t)$ is the model mismatch, and $\bs{p}$ are additional parameters in the model setup that are known, and do not need to be estimated (cf. \S\ref{s:KLbasis}, and \S\ref{s:statespacemodels} for more details). Furthermore, $\bs{u}_0(\bs{x})$ is the initial condition at time $\ti$. We presume that the state is divergence free (using the Boussinesq approximation for ABL flows), so that  $\bs{\nabla} \cdot \bs{u}_0=0$, and also $\bs{\nabla} \cdot \bs{v}=0$. Further practical details on the state equation are discussed in \S~\ref{s:statespacemodels}.

	Given a measurement series $\mathsfbi{Y}\triangleq [\bs{y}_{1}, \,\dots, \bs{y}_{N_s}]$, a related series of states $\mathsfbi{U} \triangleq [\bs{u}_{0},\,\dots, \bs{u}_{N_s}]$ is considered. Since both the measurements and the state equation contain errors ($\bs{w}_{n}$ and $\bs{v}(\bs{x},t)$ respectively) that are unknown random variables, the distributions $P(\bs{\mathsfbi{Y}})$ and $P(\mathsfbi{U})$ can be introduced. A popular approach for finding a good estimate for the state $\uvec(\bs{x},t)$ ($\ti\leq t \leq \tf$) is to find the state  $\bs{u}$ that maximises the conditional probability distribution $P(\mathsfbi{U}|\bs{\mathsfbi{Y}})$, i.e. given the measurement series  $\bs{\mathsfbi{Y}}$. Applying  Bayes' rule yields $P(\mathsfbi{U}|\bs{\mathsfbi{Y}}) = P(\bs{\mathsfbi{Y}}|\mathsfbi{U})P(\mathsfbi{U})/P(\bs{\mathsfbi{Y}})$. However, finding the state $\uvec(\bs{x},t)$ that maximises this probability, leads to an optimisation problem over the full space--time state space, that is very high dimensional. Moreover, expressing $P(\mathsfbi{U}) = P(\bs{u}_0,\ldots,\bs{u}_{N_s})$ requires knowledge of the space--time correlation function of the bias $\bs{v}(\bs{x},t)$, which is usually not straightforward.
	
	Therefore, in an alternative approach, the modeled space--time velocity field $\ft{\uvec}(\bs{x},t)$ is considered, following from
	\begin{subequations}
		\begin{align}
		\frac{\partial \ft{\bs{u}}}{\partial t} &= \bs{f}(\ft{\bs{u}},\bs{p}),\label{eq:modelbis} \\
		\ft{\uvec}(\bs{x},t_0) &= \bs{u}_0(\bs{x}). \label{eq:initialbis}
		\end{align}
	\end{subequations}
	Since this is a deterministic equation, we also introduce the solution operator $\mathcal{M}_{t}(\bs{u}_0(\bs{x}))\triangleq \ft{\uvec}(\bs{x},t)$. Further, since $\bs{h}$ is a linear function, 
	\begin{align}
	\bs{y}_n & = \bs{h}(\mathcal{M}_{t}(\bs{u}_0(\bs{x})),t_n)+ \bs{h}(\bs{\epsilon}(\bs{x},t),t_n) +\bs{w}_n, \label{eq:perfectmodelmeasurement}
	\end{align}
	with $\bs{\epsilon} \triangleq \bs{u} - \ft{\bs{u}} = \bs{u} -  \mathcal{M}_{t}(\bs{u}_0)$. Note that for linear systems, the probability distribution of $\bs{\epsilon}$ is simply a linear transformation of the distribution of $\bs{v}$. However, since this distribution is not known, we simply consider the distribution of $\bs{\zeta}_n \triangleq\bs{h}(\bs{\epsilon}(\bs{x},t),t_n) +\bs{w}_n$, with random variables $\bs{\zeta}_n$ that are independent and Gaussian distributed with same variance $\gamma^2$, which is further tuned during the setup of the optimisation problem (cf. below). We then consider the state that maximises the conditional probability $P(\ft{\mathsfbi{U}}|\bs{\mathsfbi{Y}})$. Elaborating leads to $P(\ft{\mathsfbi{U}}|\bs{\mathsfbi{Y}}) = P(\ft{\mathsfbi{U}}) P(\bs{\mathsfbi{Y}}|\ft{\mathsfbi{U}})/P(\bs{\mathsfbi{Y}}) \sim P(\bs{u}_0) \prod_{n} P(\bs{y}_n|\mathcal{M}_{t_n}(\bs{u}_0))$. Thus, using the assumed distribution for $\zeta$
	\begin{equation} 
	P(\ft{\mathsfbi{U}}|\bs{\mathsfbi{Y}}) \sim P(\bs{u}_0) \prod_{n=1}^{N_{\mt{s}}} \exp\left(\frac{\|\bs{y}_n-\bs{h}_n(\mathcal{M}_{t}(\bs{u}_0), t_n) \|^2}{2\gamma^2}\right).
	\end{equation}
	
	In order to further elaborate $P(\bs{u}_0)$, and avoid mathematical technicalities related to probability distributions over infinite dimensional spaces, it is assumed here that this distribution is finite-dimensional or can be approximated by a finte-dimensional random process. For instance, given the mean of the velocity field $\bs{U}(\bs{x})=\mean{\uvec(\bs{x})}$, and its two-point covariance $\mathsfi{B}_{ij}(\bs{x},\breve{\bs{x}})= \mean{u_i'(\bs{x}) u_j'(\breve{\bs{x}})}$ (with $\uvec'=\uvec-\bs{U}$), a truncated Karhunen--Loève decomposition can be used to arrive at a finite dimensional representation. This leads to \citep{berkooz1993proper}
	\begin{align} \label{eq:KLdecomp}
	\uvec &\approx  \bs{U}(\bs{x}) + \sum_{k=1}^{N_{\mt{m}}}a_{k}(t) \lambda_{k}^{1/2}\bs{\psi}_{k}(\bs{x})=\bs{U} + \bs{\Psi}\bs{\Lambda}^{1/2} \bs{a}, 
	\end{align}
	where $\bs{a}=[a_1, \,\ldots, a_{N_{\mt{m}}}]^\top$ are now uncorrelated random variables with unit variance, and where $\lambda_k$ are the eigenvalues of $\mathsfi{B}_{ij}$ (ordered such that $\lambda_1\geq\lambda_2\geq\cdots$), with $\bs{\Lambda} = \mt{diag}(\lambda_1, \,\ldots, \lambda_{N_{\mt{m}}})$ and $\bs{\psi}_k$ are the eigenvectors of $\mathsfi{B}_{ij}$ with $\bs{\Psi}= [\bs{\psi}_1, \,\ldots, \bs{\psi}_{N_{\mt{m}}}]$. They are obtained by solving the following Fredholm eigenvalue problem  
	\begin{equation} \label{eq:eigenvalue_int}
	\frac{1}{|\Omega|}\int_{\Omega}\mathsfbi{B}(\bs{x},\breve{\bs{x}})\bs{\psi}_{k}(\breve{\bs{x}}) \ \mathrm{d}\breve{\bs{x}} = \lambda_{k} \bs{\psi}_{k}(\bs{x}), 
	\end{equation}
	where the eigenfunctions are normalised such that $\|\bs{\psi}_{k}(\bs{x})\| = 1$. Finally, using~(\ref{eq:KLdecomp}), this leads to $P(\bs{u}_0)\sim P(\bs{a})$. Moreover, although velocity fluctuations in  turbulent boundary layers are known to deviate from a Gaussian distribution \citep{meneveau2013generalized}, it is a valid first order approximation, and thus, $P(\bs{u}_0)\sim P(\bs{a}) \sim \exp(\frac{1}{2}\|\bs{a} \|^2)$. 
	
	Bringing all above assumptions together, and formulating the optimisation in terms of $\log[P(\ft{\mathsfbi{U}}|\bs{\mathsfbi{Y}})]$, which does not change the optimum, leads to \citep[see e.g. ][]{jazwinski2007stochastic}
	\begin{align}
	&\underset{\bs{a}}{\mt{minimise}}  &&\J(\bs{a}) =  \frac{1}{2}\|\bs{a}\|^2 +  \frac{1}{2\gamma^2} \sum_{n=1}^{N_{\mt{m}}} \left\|\bs{y}_n-\bs{h}\left(\mathcal{M}_{t}(\bs{U} + \bs{\Psi}\bs{\Lambda}^{1/2} \bs{a}), t_n\right) \right\|^2. \label{eq:optprob}
	\end{align}
	We note that the value of $\gamma^2$ is at this point unknown -- further selection is discussed in \S~\ref{ss:lambdatuning}. Above problem can be interpreted as minimizing the model--measurement mismatch, given a regularisation  $\frac{1}{2}\|\bs{a}\|^2$, that is particularly important in areas where no measurement information is given. The formulation based on a Karhunen--Loève decomposition as discussed above, leads to a regularisation that is optimally conditioned. If the velocity field $\bs{u}_0$ were to be used directly as optimisation variable, the regularisation would typically take the form $\frac{1}{2}\|\bs{u}_0\|^2_{\mathsfbi{B}^{-1}}$, with a condition number that corresponds to $\lambda_1/\lambda_N$ (given $N$ degrees of freedom in the finite-dimensional representation of $\bs{u}_0$).

	\subsection{Formulation of an efficient Karhunen--Loève basis}\label{s:KLbasis}
	In order to use the approach proposed in \S\ref{s:general_methodology}, a Karhunen--Loève basis for the wind-field distribution is required. This can in principle be based on an ensemble of all possible wind fields that occur in the ABL assembled over many years. However, acquiring the covariance tensor for this would be nontrivial, and the resulting basis may require many modes for an accurate parameterisation of all possible states. Therefore, we envisage a different approach, in which the covariance tensor is parametrised depending on a number of background parameters $\bs{p}$ (e.g. wind direction, friction velocity, surface roughness, Obukhov length, Rossby number, etc.). These parameters are presumed to change slowly in time, and are either given, or estimated using a different overarching algorithm. 
	
	In the current manuscript, we focus in particular on neutral conditions, and simplify the approach to that of estimating the wind field in a neutral pressure driven boundary layer in equilibrium (the extension to Ekman layers should be straightforward, but is not considered here). For this case, the relevant background parameters are given by $\bs{p}=[u_*, z_0, H, \theta]$, with $u_*$ the friction velocity, $z_0$ the surface roughness, $H$ the boundary layer height, and $\theta$ the mean wind direction. The mean-velocity profile in is then given by \citep[see e.g. ][]{pope2000turbulent}
	\begin{equation}
	\bs{U}(\bs{x}) = \frac{u_*}{\kappa}\left(\log\left(\frac{z}{z_0}\right) + F\left(\frac{z}{H}\right)\right)\mathsfbi{Q}_3\bs{e}_1, 
	\end{equation} 
	with $z\triangleq x_3$ and $F(z/H)$ an outer-layer velocity deficit function, that in the classical picture of inner--outer separation, can be uniquely determined from a single simulation. Further, $\mathsfbi{Q}_3(\theta)$ is a rotation matrix around around the wall normal direction that reorients the coordinate axis in the horizontal plane to a system with main wind direction $\theta$. Similarly, using outer-layer scaling arguments \citep{townsend1980structure}, and horizontal homogeneity, the covariance tensor can be expressed as
	\begin{equation}
	\mathsfbi{B}(\bs{x},\breve{\bs{x}}) = u_*^2 \ \mathsfbi{Q}_3 \mathsfbi{B}^+\left(\frac{x_1-\breve{x}_1}{H},\frac{x_2-\breve{x}_2}{H},\frac{x_3}{H},\frac{\breve{x}_3}{H}\right)\mathsfbi{Q}_3^\top ,
	\end{equation}
	with $\mathsfbi{B}^+$ the covariance normalised by friction velocity and boundary layer height, obtained in a system with main wind direction oriented in the $x_1$ direction. We note that as a result of outer-layer similarity, $\bs{B}$ will not depend on the surface roughness $z_0$ \citep[see e.g. ][]{squire2016comparison}.\

	It is easily shown that for horizontal homogeneous directions, as encountered approximately in atmospheric boundary layer flows, eigenfunctions simply correspond to Fourier modes \citep{berkooz1993proper}, such that $\bs{\psi}_{\bs{k},m}(\bs{x}) = \widehat{\bs{\psi}}_{\bs{k},m}(x_3)\exp(\mt{i}(k_1x_1+k_2x_2))$. Inserting this expression in (\ref{eq:eigenvalue_int}) and integrating over $x_1$ and $x_2$ gives
	\begin{equation}
	\frac{1}{H}\int_{0}^{H}\widehat{\mathsfbi{B}}(\bs{k},x_3,\breve{x}_3)\widehat{\bs{\psi}}_{\bs{k},m}(\breve{x}_3)\ \mathrm{d}\breve{x}_3 = \widehat{\lambda}_{\bs{k}, m} \widehat{\bs{\psi}}_{\bs{k},m}(x_3), \label{eq:homogeneousPODproblem}
	\end{equation}
	normalised such that $\|\widehat{\bs{\psi}}_{\bs{k},m}\|=1$, where $\widehat{\mathsfi{B}}_{ij}(\bs{k},x_3,\breve{x}_3) = \langle \widehat{u}_i(\bs{k},x_3)\widehat{u}_j^*(\bs{k},\breve{x}_3) \rangle$ is the horizontal Fourier transform of the correlation tensor, with $\bs{k}=[k_1,k_2]^{\top}$.

	In order to further elaborate, consider velocity fields that are discretised in space on a domain $L_1\times L_2 \times H$. Consistent with our LES code (cf. \S\ref{s:statespacemodels}), we consider $N_1 \times N_2 \times N_3$ grid points that are uniformly distributed, and a grid that is staggered in the vertical direction for the $u_3$ component, leading to $N_1N_2(3N_3-1)$ degrees of freedom. Thus, wavenumbers in horizontal directions are integer multiples of $k_1^*\triangleq2\upi/L_1$, $k_2^*\triangleq2\upi/L_2$, with cutoff the wave number corresponding to $\bs{k}_c\triangleq[\upi/\Delta_1, \upi/\Delta_2]^\top$, with $\Delta_1=L_1/N_1$, and $\Delta_2=L_2/N_2$.  Given this setup, (\ref{eq:homogeneousPODproblem}) requires the solution of $\frac{1}{4}N_1N_2$ eigenvalue problems (factor $\frac{1}{4}$ because of symmetries), each of size $(3N_3-1)\times(3N_3-1)$. This allows an easier construction of higher dimensional basis, compared to, e.g., the snapshot POD method \citep{sirovich1987turbulence}, where the rank is limited to the amount of samples used for the determination of $\widehat{\mathsfbi{B}}$. 
	
	Expanding the velocity field using the POD eigenfunctions leads to  
	\begin{align} \label{eq:realeigenfunction}
	\bs{u}'&=\sum_{\bs{k},m}c_{\bs{k},m}\widehat{\lambda}^{1/2}_{\bs{k}, m}\bs{\psi}_{\bs{k},m} \\
	&=\sum_{m}c_{\bs{0}, m}\widehat{\lambda}^{1/2}_{\bs{k}, m}\bs{\psi}_{\bs{0},m} + 2\sum_{\bs{k}^+,m}\widehat{\lambda}^{1/2}_{\bs{k}, m}\left(\Real(c_{\bs{k}, m})\Real(\bs{\psi}_{\bs{k},m}) -\Imag(c_{\bs{k}, m})\Imag(\bs{\psi}_{\bs{k},m})\right), \nonumber
	\end{align}
	where in the second step the conjugate symmetry of $\widehat{\mathsfbi{B}}$ is used. This transforms the complex modes $\bs{\psi}_{\bs{k},m}$ to an equivalent set of real orthogonal modes, $\Real(\bs{\psi}_{\bs{k},m})$ and $\Imag(\bs{\psi}_{\bs{k},m})$, by adding together positive and corresponding negative wave numbers. In this way conjugate symmetry constraints in the optimisation problem are avoided. The wavenumber $\bs{k}^+$ is chosen such that all coefficients are only used once, here we use $\bs{k}^+= \bs{k}\mid (k_1>0~\text{or}~k_1=0, k_2>0)$. For $\bs{k}=\bs{0}$ the modes are real, and therefore the complex coefficients are omitted. Grouping all coefficients, eigenvalues and modes gives
	\begin{subequations}
		\begin{align}
		\widehat{\bs{a}} &= \begin{bmatrix}c_{\bs{0},1} & \cdots& \sqrt{2}\Real(c_{\bs{k}_c,3N_z})& \sqrt{2}\Imag(c_{\bs{k}_c,3N_z})\end{bmatrix}, \\
		\widehat{\bs{\Lambda}} &= \mt{diag}\begin{pmatrix}\lambda_{\bs{0},1} & \cdots&\lambda_{\bs{k}_c,3N_z}& \lambda_{\bs{k}_c,3N_z}\end{pmatrix}, \\
		\widehat{\bs{\Psi}}&=\begin{bmatrix}\bs{\psi}_{\bs{0},1}&  \cdots& \sqrt{2}\Real(\bs{\psi}_{\bs{k}_c,3N_z})& -\sqrt{2}\Imag(\bs{\psi}_{\bs{k}_c,3N_z})\end{bmatrix},
		\end{align}
	\end{subequations}
	where the scaling factor $\sqrt{2}$ is added such that $\|\widehat{\Psi}_k \|=1$, and $\langle\widehat{a}_k^2 \rangle=1$. Converting to a basis using the $N_{\mt{m}}$ most energetic modes can be done by
	\begin{subequations}
		\begin{align}
		\bs{a} &= \bs{S}\mathsfbi{P}\widehat{\bs{a}}, \\
		\bs{\Lambda} &= \mathsfbi{S}\mathsfbi{P}\widehat{\bs{\Lambda}}\mathsfbi{P}^\top\mathsfbi{S}^\top, \\
		\bs{\Psi} &= \widehat{\bs{\Psi}}\mathsfbi{P}^\top\mathsfbi{S}^\top, 
		\end{align}
	\end{subequations}
	where $\mathsfbi{P}$ represents a permutation matrix ordering the eigenvalue in descending order, and $\mathsfbi{S} = \begin{bmatrix} \mathsfbi{I} & \bs{0} \end{bmatrix}$ is a selection matrix that removes all modes with order higher than $ N_{\mt{m}}$, and $\mathsfbi{I}$ is an identity matrix of size $N_{\mt{m}}\times N_{\mt{m}}$. We note that it is required to select $N_m\leq N_1N_2(2N_3-1)+1$. This results from the fact that the velocity field is solenoidal. In the discretised LES system (cf. next section), there are $N_1N_2N_3-1$ independent continuity constraints, and this leads to a subspace of $N_1N_2N_3-1$ modes in the discrete POD basis that are orthogonal to the solenoidal subspace, with eigenvalue zero.

	\section{Description of the state space models}\label{s:statespacemodels}
	\subsection{Large-eddy simulations}
	The main state-space model that we consider in the current work is based on large-eddy simulations of a neutrally stable pressure driven boundary layer. We presume knowledge of the setup parameters $\bs{p}=[u_*, z_0, H, \theta]$, and simply orient the simulation domain such that $\bs{e}_1=\bs{e}_\theta$. The governing equations then correspond to 
	\begin{subequations}
		\begin{align}
		\frac{ \partial \bs{\uvecfi}}{\partial t} &= -\uvecfi\cdot\bs{\nabla} \bs{\uvecfi}  + u_*^2 H \bs{e}_1 - \frac{1}{\rho} \bs{\nabla}  \ft{p}  + \bs{\nabla}\cdot\bs{\tau}_{\mt{SGS}}, \label{eq:NS_mom} \\
		0 &=\bs{\nabla} \cdot\bs{\uvecfi} , \label{eq:NS_con}
		\end{align}
	\end{subequations}
	where $\uvecfi$ is the filtered velocity field,  $-\rho u_*^2 H \bs{e}_1$ is the mean background pressure gradient, and $\ft{p}$ is the remaining filtered pressure fluctuation. For the subgrid-scale model $\bs{\tau}_{\mt{SGS}}$ we use the Smagorinsky model \citep{smagorinsky1963general} with Smagorinsky coefficient $C_s=0.14$ in combination with wall-damping \citep{mason1992stochastic} with $n=1$. 
	
	In horizontal directions, we consider a computational domain that is sufficiently large for boundary conditions not to influence the solution in the estimation region of interest, so that simple periodic boundary conditions can be used, and boundary fields in space do not need to be estimated. In the vertical direction, nonpermeable slip boundary conditions are used both along top and bottom wall. Along the bottom wall, this is supplemented with a classical wall model \citep{moeng1984large}, following the implementation proposed by \citet{bou2005scale}; see also \citet{meyers2011error} for further details. 
	
	All state-space simulations are performed using our in-house LES code SP-Wind \citep{meyers2007,munters2016shifted}, in which above model is implemented using a pseudospectral discretisation in the horizontal directions, and a fourth-order energy conserving scheme in the vertical directions \citep{verstappen2003symmetry}. For the time integration we use a fourth-order Runge--Kutta method, where the time step is fixed, approximately corresponding to a CFL number of 0.4.
	
	\subsection{Taylor's-frozen turbulence model}
	As an alternative simpler reference, we also consider the Taylor's frozen turbulence (TFT) model  \citep{taylor1938spectrum} as a state-space model. It corresponds to $\uvecfi(\bs{x},t)=\mathcal{M}_{t}(\bs{u}_0(\bs{x}))=\bs{u}_0(\bs{x}-U_{\infty}\bs{e}_1(t-\ti))$, with $U_{\infty}$ a characteristic convection speed, or equivalently in differential form  
	\begin{align}
	\frac{\partial \uvecfi}{\partial t} = -U_{\infty}\bs{e}_1\cdot \bs{\nabla} \uvecfi.
	\end{align}
	For $U_{\infty}$, we use the lidar mount-height velocity, which is simply estimated as $U_{\infty}/u_*= \kappa^{-1}\log(z_{\mt{m}}/z_{0})$. The model is solved using the same time integration and spatial discretisation scheme as the LES code. Only horizontal boundary conditions are necessary, for which we also use periodicity.

	\section{Optimisation methodology and adjoint equations} \label{sec:Opt}
	For the optimisation problem (\ref{eq:optprob}) we use the L-BFGS algorithm from \citep{byrd1995limited}, which is a quasi-Newton algorithm suitable for large scale problems. The step length is determined by the Moré--Thuente line-search algorithm \citep{more1994line}. Due to the vast amount of optimisation variables, finite difference gradient calculation is computationally not feasible, and instead a continuous adjoint approach is applied. 
	
	The sensitivity of the cost function is given by 
	\begin{align}
	\frac{\partial \J}{\partial a_i}= a_i -\lambda_i^{1/2}\int_{\Omega}\bs{\psi}_i(\bs{x})\cdot\uvecfia(\bs{x},\ti)\mathrm{d}\bs{x},
	\end{align} 
	where $\uvecfia$ is the adjoint variable (see Appendix $\ref{sec:adjoint_deriv}$ for a derivation), which follows from solving the adjoint equations (cf. below). In this way, the gradient can be computed to all the variables at the cost of one extra simulation, which has a similar form and computational cost as the forward equations.  
	
	The derivation of the continuous adjoint equations is quite standard, and not repeated here (see  \citet{bewley2001dns, goit2015optimal} for respectively a DNS derivation, and a derivation of the additional LES specific terms). The adjoint equations can be summarised as
	\begin{subequations} \label{eq:adjmom}
		\begin{align}
		-\frac{\partial \uvecfia}{\partial t} &= \uvecfi\cdot \bs{\nabla}\uvecfia+\uvecfia\cdot\bs{\nabla}\uvecfi+\bs{\nabla}\cdot\bs{\tau}^*_{SGS} + \frac{1}{\rho}\bs{\nabla} \pfila + \sum_{i=1}^{N_{\mt{r}}}\bs{f}_{i},  \\
		\bs{\nabla} \cdot \uvecfia &= 0.
		\end{align}
	\end{subequations}
	Here $\uvecfia$ and $\pfila$ are respectively the adjoint velocity and pressure variable, $\bs{\tau}^*_{SGS}$ is the adjoint subgrid-scale model, and $\bs{f}_{i}$ is the adjoint forcing term connected to measurement $i$ (see Appendix $\ref{sec:adjoint_deriv}$ for a derivation), and given by
	\begin{align} \label{eq:adjforcing}
	\bs{f}_{i} = \frac{1}{\gamma^2T_{\mt{s}}}\sum_{n=1}^{N_{\mt{s}}}(y_{n,i}-h_{n,i}) \, \mathcal{G}_{\mt{l}}\!\left(\mathsfbi{Q}(t)(\bs{x}-\bs{x}_{i}(t))\right)\bs{e}_{\mt{l}}(t) \, \mt{H}\!\left(\frac{T_{\mt{s}}}{2}-\left|t - t_{n-\frac{1}{2}}\right|\right),
	\end{align}
	where $\mt{H}$ is the Heaviside function. The adjoint equations need to be integrated backwards in time. The starting condition for the adjoint variables specified at the end of the horizon is simply $\uvecfia(\bs{x},\tf) = 0$ (see Appendix \ref{a:PODconstraint} for details). For the integration we use a fourth-order discrete adjoint Runge--Kutta scheme \citep{hager2000runge}, which is consistent with our forward time integration. The spatial boundary conditions correspond to periodicity in the horizontal directions, and impermeability for the bottom and top of the domain in combination with respectively an adjoint wall model at the bottom. Details are found in \citet{goit2015optimal}.

	The adjoint TFT equations are very similar to the adjoint LES-equations, i.e.
	\begin{align}
	-\frac{\partial \uvecfia}{\partial t} &= \bs{U}_{\infty}\cdot \bs{\nabla}\uvecfia + \sum_{i=1}^{N_{\mt{r}}}\bs{f}_{i}.
	\end{align}
	These equations also have to be solved backwards in time with initial condition $\uvecfia_0 = 0$ and periodic boundary conditions in space. The forcing term $\bs{f}_{i}$ is the same as the one used by the adjoint LES (\ref{eq:adjforcing}).
	
	\section{Case description} \label{sec:case_setup}
	\subsection{Simulation setup}
	\begin{figure} 
		\includegraphics[width=\textwidth]{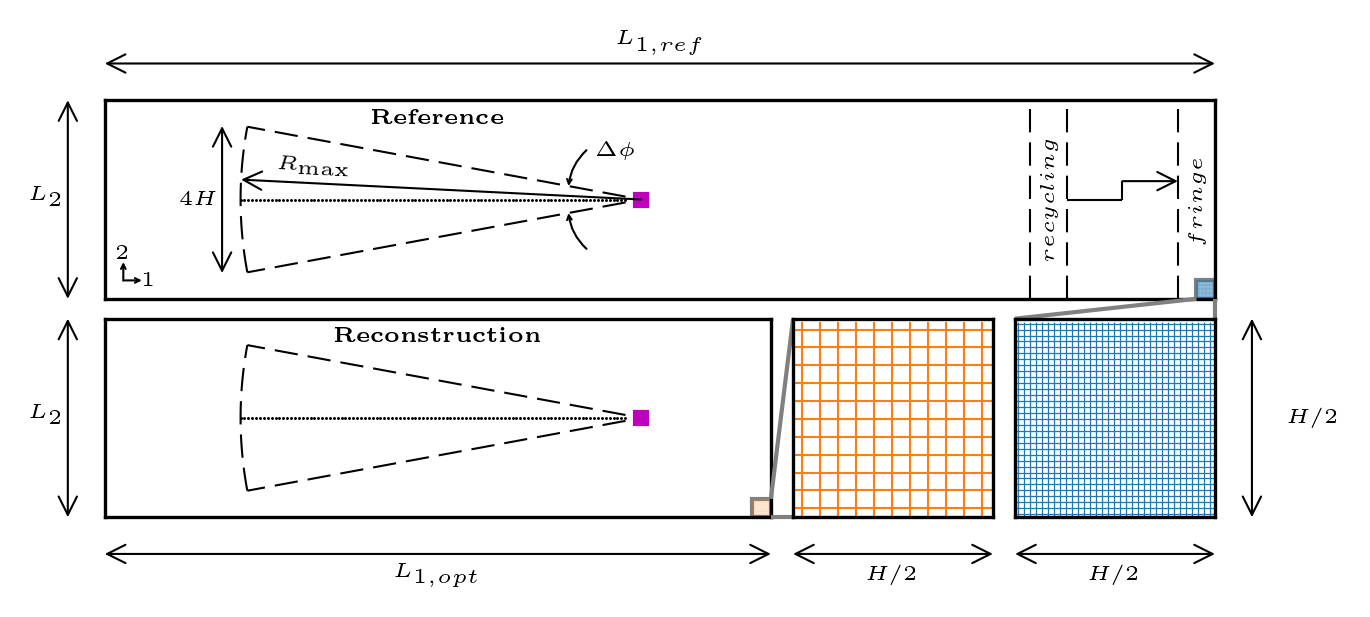}%
		\caption{Schematic planview of the different simulation domains used in this study. The top and bottom domain respectively represent the reference and reconstruction domain. The lidar mount location (purple rectangle) and the scanned area  (in PPI mode) are shown in dashed lines. For the reference domain the location of the recycling and fringe region used for the implementation of shifted-periodic boundary conditions, are also shown. }
		\label{fig:simulation_overview}
	\end{figure}
	
	In the current work, a reference simulation is used to take virtual lidar measurements.  Afterwards the reconstruction of the flow field is performed on a different domain with a coarser mesh. Both domains are schematically represented on figure \ref{fig:simulation_overview}, and the details are summarised in table \ref{tab:SimSetup} and further discussed below. All simulations use a boundary layer height of $H=\SI{1}{\kilo\metre}$ and a surface roughness of $z_{0}=\SI{0.1}{\metre}$, which is a common overland value, and a friction number of $u_* = \SI{0.5}{\metre \second^{-1}}$, which leads to a wind speed of around $\SI{8}{\metre \second^{-1}}$ at $\SI{100}{\metre}$.
	
	For the reference simulation we use a relatively fine grid resolution of $0.015H \times 0.015H\times 0.005H$, combined with a long domain $30H \times 5.4H\times H$ to avoid spurious periodic correlations. In addition, the boundary conditions of the reference domain are shifted over $d_s=H$ to ensure statistically homogeneous inflow, and avoid locking of large-scale motions \citep[see Ref.][for details]{munters2016shifted}. The fringe region is located between $15.5H$ and $14.5H$ upstream of the lidar mount, the distance between recycling and fringe region is chosen as $3H$, this is also visualised on figure \ref{fig:simulation_overview}. A spinup period of $50H/u_*$ is used, to ensure a statistical steady state has been reached, before virtual lidar measurements are taken. 
	
	The optimisation domain is smaller than the reference domain, but should at least encompass the region of influence of the time series of lidar measurements. Based on the lidar setup discussed in \S\ref{ss:scanning_trajectories} and the assimilation time horizon of $0.1H/u_*$ (see \S\ref{ss:optimization}), suitable domain dimensions are found to be $18H \times 5.4H\times H$. Reconstruction of velocity scales smaller than the lidar filter size is not possible, and therefore adding these scales is not improving the reconstruction, and needlessly increases the computational complexity. Therefore the grid is chosen to be $0.05H \times 0.05H\times 0.0167H$. Note that by using a different grid resolution for reference and reconstruction, a bias is introduced in the state space model, although admittedly, the expected bias with respect to real measurements can be quite different (the use of real data is however not in the scope of the current study).
	
	The two-point covariance matrix $\bs{B}$ is found to be sensitive to the grid resolution and is therefore computed on the same grid used for the reference simulation. The domain is chosen as $18H \times 5.4H\times H$ -- a tradeoff between accuracy and reasonable computational cost, caused by the long time averaging that is necessary to get sufficient statistical convergence (see further). We note that the largest flow structures in an ABL, i.e., streamwise streaks, are of the same scale as our domain. Therefore, we find that the far correlations of the streamwise velocity component are influenced by the periodic boundary conditions. This effect is briefly discussed in \S \ref{sec:covariance}. The simulation is spun up over a period of $50H/u_*$, and subsequentially sampled every $0.01H/u_*$ over a time horizon of $100H/u_*$, leading to $10^4$ samples. Note that the equations are reflection symmetric with respect to a streamwise-vertical plane, and therefore, we add the mirrored samples as well. Finally, the two-point covariance tensor needs to be converted from the correlation to the optimisation domain. The domain dimensions are the same, such that only a restriction to the coarser grid is needed. For the horizontal and vertical directions, we use a cutoff filter and linear interpolation respectively. To assure that the POD modes obtained from the restricted covariance matrix are solenoidal, the restricted matrix is projected on the solenoidal subspace of the coarser grid by means of a Helmholtz decomposition.
	
	\subsection{Lidar setup} \label{ss:scanning_trajectories}
	We demonstrate the methodology using two scanning modes. To keep the two trajectories comparable, we use the same scanning period of $T_{\mt{p}}=0.1H/u_*=\SI{200}{\second}$. The lidar mount is located at $\bs{x}_{\mt{m}}=[0, 0, 0.1H]$ for both cases.

	In a first case study, we set the virtual lidar in plan position indicator (PPI) scanning mode with zero elevation angle, thus tracking a horizontal sweeping trajectory. The direction of the beam is given by $\bs{e}_{\mt{l}}(t) = \mathsfbi{Q}_3(\phi(t))\bs{e}_1$, such that $\mathsfbi{Q}(t)=\mathsfbi{Q}_3(t)$. The azimuthal angle $\phi$ is given by $\phi(t) = \Delta \phi\text{ Triag}(t/T_{\mt{p}}) + \upi$, where $\mt{Triag}(t)\triangleq 1/\upi\, \mt{sin}^{-1}(\mt{sin}(2\upi t))$ is the triangle wave function with unit amplitude and period, such that the lidar has a constant azimuthal angular velocity of $|\partial \phi/\partial t|=2\Delta \phi /T_{\mt{p}}$. For the azimuthal range $\Delta \phi$ we take  $\Delta \phi = 2\sin^{-1}(2H/R_{\mt{max}})$ (see figure \ref{fig:simulation_overview}).
	
	In a second case study, we study a 3D scanning pattern. First we define a parametric curve $\bs{l}(t)$, where we use $\bs{l}(t)=[1, Asin(\omega_2 t -\delta),B(sin(\omega_3t)+1)]$, a special case of a 3D Lissajous curve, in which $A$ and $B$ respectively control  the horizontal and vertical extent of the trajectory, and the ratio of angular speeds $\omega_3/\omega_2$, and phase $\delta$ control the shape of the curve. The construction is such that the lidar does not scan lower than the lidar mount height. The lidar direction is given by $\bs{e}_{\mt{l}}(t) = \mathsfbi{Q}_3(\upi)\bs{l}(t)/||\bs{l}(t)||$. For this study we respectively take $A=\tan(\frac{1}{2}\Delta \phi)$, which is easily verified to lead to the same spanwise extent as the sweeping lidar, and $B=\tan\Delta \theta$ with $\Delta \theta = \sin^{-1}(0.8H/R_{\mt{max}})$, which gives a maximum scanning altitude of $0.9H$. Further, $\omega_2=4\upi/T_{\mt{p}}$, $\omega_3=3/2\omega_2$, and $\delta=\upi/2$.

	\begin{table}
		\centering
		\begin{tabular}{l l l c}
			\textit{Reference domain}\Tstrut    \span \span                \\
			Domain size\Tstrut      &$L_1\times L_2  \times H$  &$\SI{30}{km}\times \SI{5.4}{km}\times\SI{1}{km}$  \\
			Grid size & $N_1\times N_2 \times N_3$& $2000 \times 360 \times 200 $\\
			Cell size & $\Delta x_1 \times\Delta x_2 \times\Delta x_3 $& $\SI{15}{m}\times\SI{15}{m}\times\SI{5}{m}$ \\
			\textit{Correlation domain}\Tstrut    \span \span                \\
			Domain size\Tstrut      &$L_1\times L_2  \times H$  &$\SI{18}{km}\times \SI{5.4}{km}\times\SI{1}{km}$  \\
			Grid size & $N_1\times N_2 \times N_3$& $1200 \times 360 \times 200 $\\
			Cell size & $\Delta_1 \times\Delta_2 \times\Delta_3 $& $\SI{15}{m}\times\SI{15}{m}\times\SI{5}{m}$\\
			\textit{Reconstruction domain}\Tstrut    \span \span                \\
			Domain size\Tstrut      &$L_1\times L_2  \times H$  &$\SI{18}{km}\times \SI{5.4}{km}\times\SI{1}{km}$  \\
			Grid size & $N_1\times N_2 \times N_3$& $360 \times 108 \times 60 $\\
			Cell size & $\Delta_1 \times\Delta_2 \times\Delta_3 $& $\SI{50}{m}\times\SI{50}{m}\times\SI{16.67}{m}$\\
			\textit{General simulation parameters} \Tstrut    \span \span                \\
			Roughness length\Tstrut      &$z_{0}$ & $\SI{0.1}{m}$\\
			Friction velocity &$u_*$& $\SI{0.5}{m.s^{-1}} $ \\
		\end{tabular}
		\caption{Summary of the setup parameters of the different simulations.}
		\label{tab:SimSetup}
	\end{table}
	
	\subsection{Optimisation setup} \label{ss:optimization}
	
	\begin{table}
		\centering
		\begin{tabular}{l l l c}
			Optimisation method\Tstrut & L-BFGS-B &  \\
			Wolfe conditions & $c_1$, $c_2$ & $10^{-4}$, $0.9$  \\
			Hessian correction pairs & m & 8 \\
			BFGS iterations & $N_{\mt{it}}$ & 300 \\
			Optimisation time window & $T$ & $0.1H/u^*$ \\
		\end{tabular}
		\caption{Summary of the optimisation parameters.}
		\label{tab:OptSetup}
	\end{table}
	
	The time horizon of the optimisation $T\triangleq\tf - \ti$ is chosen equal to a single scanning period of the lidar $T_{\mt{p}}$. Very long horizons lead to very large gradients due to the chaotic behaviour of turbulence. Moreover, since we don't explicitly include model errors, and optimally match $\uvecfi$ to the measurements, $\uvecfi$ and  $\uvec$ will diverge to the point where new measurements do not contribute to the reconstruction. 
	
	All modes with positive eigenvalues are taken into the POD basis, such that we optimise over the full space of solenoidal velocity fields. The amount of modes corresponds to $N_{\mt{m}} = N_1N_2(2N_3-1)+1$, which is obtained by substracting the  number of independent continuity constraints $N_1N_2N_3-1$ from the total degrees of freedom $N_1N_2(3N_3-1)$. In table \ref{tab:OptSetup} the most important optimisation parameters are summarised. For the optimisation, we use 8 Hessian correction pairs for the L-BFGS-B method. For the Wolfe condition parameters we take $c_1=10^{-4}$ and $c_2=0.9$, which are standard values selected for quasi-Newton methods \citep{nocedal2006numerical}. In the following sections we elaborate further on the convergence and stopping criteria for the optimisation, and on the tuning of the variance $\gamma^2$ of to the model--measurement uncertainty.
	
	\subsubsection{Adjoint gradient calculation and optimisation convergence}
	
	\begin{figure} 
		\includegraphics[width=\textwidth]{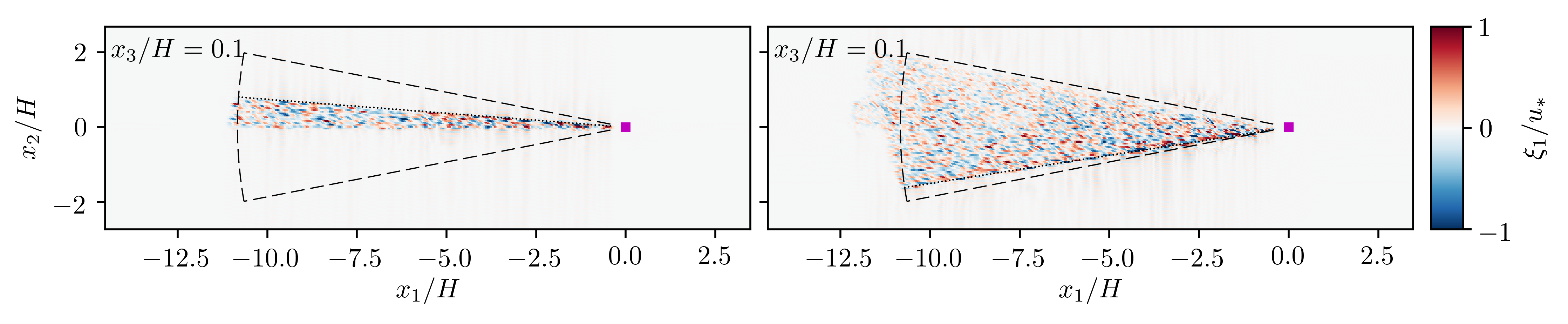}%
		\caption{The left and right hand side respectively represent the streamwise component of the adjoint velocity field $\ufia_1$ at $\tf-0.1H/u_*$ and  $\tf-0.7H/u_*$. The top and bottom represent respectively a $x_1$--$x_3$ plane cross-section at $x_2=L_2/2$ and a $x_1$--$x_2$ plane cross-section at $x_3=0.1H$. The purple square gives the lidar mount location, the dots represent the centre of all the range gates, the dashed line gives the outer edge of the scanning region. }
		\label{fig:adjointfield}
	\end{figure}
	
	In this section we discuss general properties of the optimisation. As a test case we use the PPI-scanning mode in combination with the LES reconstruction model. We use $\gamma^2=10^{-4}$ for the model--measurement uncertainty (this is further elaborated in \S \ref{ss:lambdatuning}). 
	
	First of all, the streamwise component of the adjoint field is given in figure~\ref{fig:adjointfield}. The field gives a representation of the sensitivity of the cost function to a local perturbation in velocity and the propagated effect through the Navier--Stokes equations. The adjoint field is clearly seen to originate from the lidar measurement locations, due to the forcing term of the adjoint equations being a convolution of the mismatch between the observed and simulated lidar measurements with the lidar filter kernel. The adjoint velocity field propagates upstream due to the reverse sign of the convection term. Note that for a large part of the domain, the adjoint field remains (almost) zero, which indicates that flow information in this region does not influence the measurements. The accuracy of the adjoint gradient is also verified by comparing against a finite-difference evaluation for a few selected perturbation directions. We find that the relative error of the gradient remains smaller than $0.1\%$ for the selected cases. More details are provided in appendix \ref{sec:adjoint_valid}.

	\begin{figure} 
		\includegraphics[width=\textwidth]{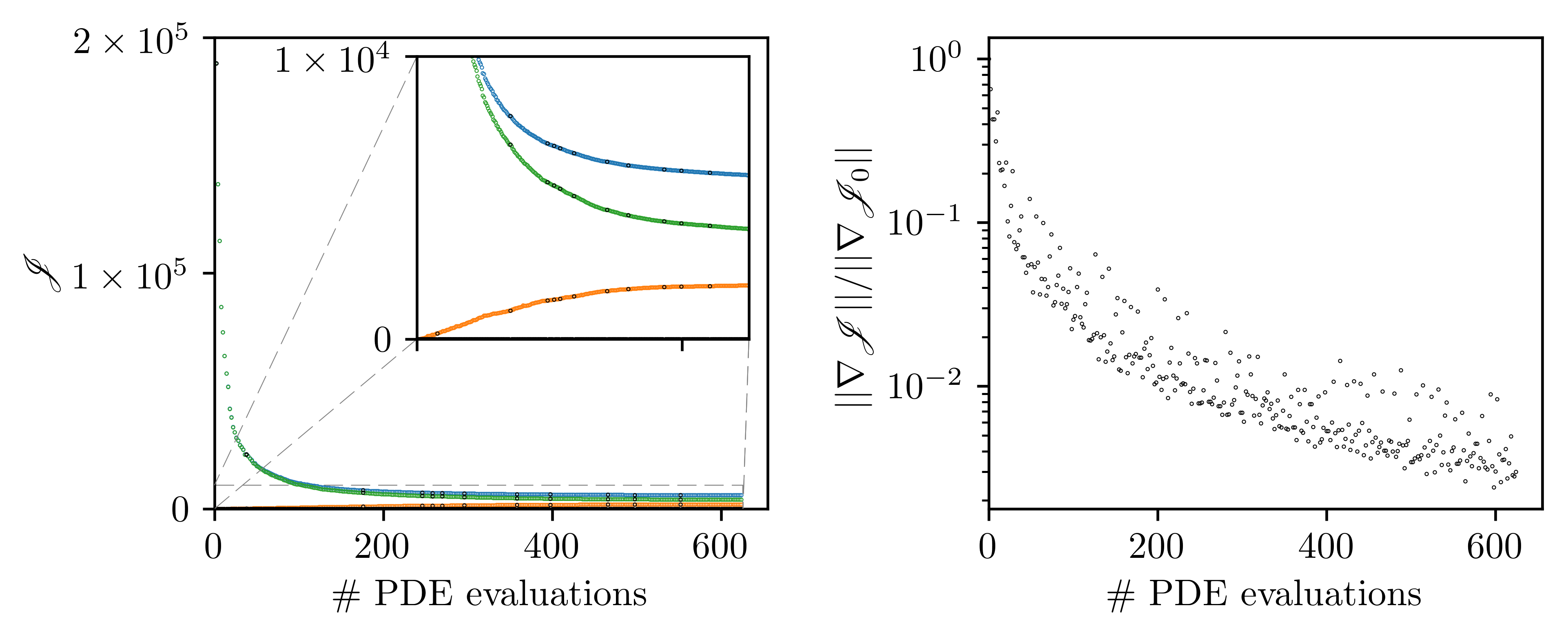}%
		\caption{The right and left figure respectively represent the reduction of the cost function and gradient as a function of the amount of PDE evaluations. The orange, green and blue dots respectively represent the $\J_{\mt{o}}$, $\J_{\mt{b}}$ contribution of the cost function and the whole cost function $\J = \J_{\mt{o}} + \J_{\mt{b}}$.}
		\label{fig:conv_opt}
	\end{figure}
	
	All optimisations are started from an initial guess $\bs{a} = \bs{0}$. The convergence history of the PPI case with $\gamma^2=10^{-4}$ is shown in figure~\ref{fig:conv_opt}. The left panel of the figure shows the evolution of the cost function as a function of the number of PDE evaluations (LES or adjoint) during the optimisation. For sake of analysis, we split the cost function (see (\ref{eq:optprob})) into two parts, i.e. $\J = \J_{\mt{b}} + \J_{\mt{o}} $, with
	\begin{align}
	\J_{\mt{b}} = \frac{1}{2}\|\bs{a}\|^2, ~~~
	\J_{\mt{o}} = \frac{1}{2\gamma^2} \sum_{n=1}^{N_{\mt{m}}} \left\|\bs{y}_n-\bs{h}\left(\mathcal{M}_{t}(\bs{U} + \bs{\Psi}\bs{\Lambda}^{1/2} \bs{a}), t_n\right) \right\|^2.
	\end{align} 
	The first represents the background variability (of the initial condition), and acts as a regularisation term, while the second represents the variability of the observation--model mismatch. Both are shown in the convergence history in figure~\ref{fig:conv_opt} as well.
	
	Each outer optimisation iteration requires a simulation of the forward and adjoint equations, and an additional line search in case the Wolfe conditions are not satisfied. The latter is found to only happen in 15 occasions, so that the number of total PDE simulations (shown in the horizontal axes in figure~\ref{fig:conv_opt}) is by good approximation twice the number of iterations. The right panel of figure~\ref{fig:conv_opt} shows the $L_2$-norm of the gradient vector $\nabla_{\bs{a}} \J$. Since the optimisation problem is unconstrained, a (local) optimum will correspond with $\nabla_{\bs{a}}\J=\bs{0}$. Given the accuracy of the gradient, which is found to be $\textit{O}(10^{-3})$ (see appendix \ref{sec:adjoint_valid}), we see that the gradient converges up to a relative value of $2\times 10^{-3}$. 
	
	\subsubsection{Tuning of the combined model--measurement uncertainty} \label{ss:lambdatuning}
\begin{figure} 
		\centering
		\includegraphics[width=0.7\textwidth]{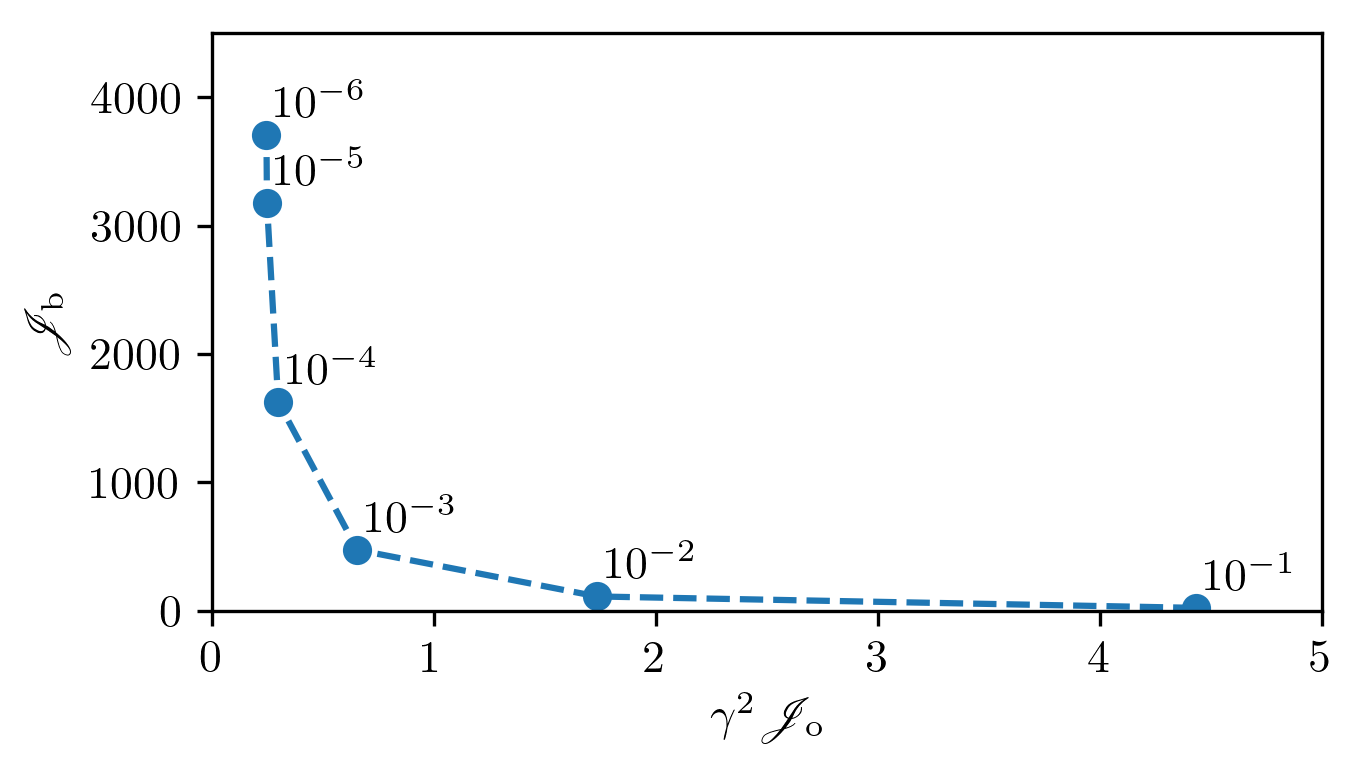}%
		\caption{Pareto front for the optimisation using the LES model, between the the regularisation part of the cost function $\J_{\mt{b}}$ and the model--observation mismatch $\gamma^2\J_{\mt{o}}$. The annotations at the markers denote different variances $\gamma^2$ of the combined model and measurement uncertainty. }
		\label{fig:pareto}
	\end{figure}

	In this section the unknown variance $\gamma^2$, introduced in \S \ref{s:general_methodology}, is tuned. The regularisation term of the cost function  $\J_{\mt{b}}$ controls the complexity of the solution, while $\gamma^2 \J_{\mt{o}}$ represents the differences between the simulated and observed measurements. By controlling the parameter $\gamma^2$, the relative trust in the model--measurement accuracy is adjusted. For the tuning, we consider the PPI scanning mode combined with the LES model. In figure \ref{fig:pareto} we show the Pareto front of $\J_{\mt{b}}$ versus $\gamma^2\J_{\mt{o}}$, for different values of $\gamma^2$. It is clearly seen that below $\gamma^2=10^{-4}$ there is almost no decrease in $\gamma^2\J_{\mt{o}}$ while the complexity of the solution increases significantly, which is an indication of overfitting. Therefore we use $\gamma^2=10^{-4}$. This value is also used for the TFT model and for the Lissajous scanning mode. 

	\section{Results} \label{sec:results}
	\subsection{Covariance matrix and POD modes} \label{sec:covariance} 
	
	\begin{figure}
		\centering
		\begin{subfigure}[b]{\textwidth}
			\includegraphics[width=1\linewidth]{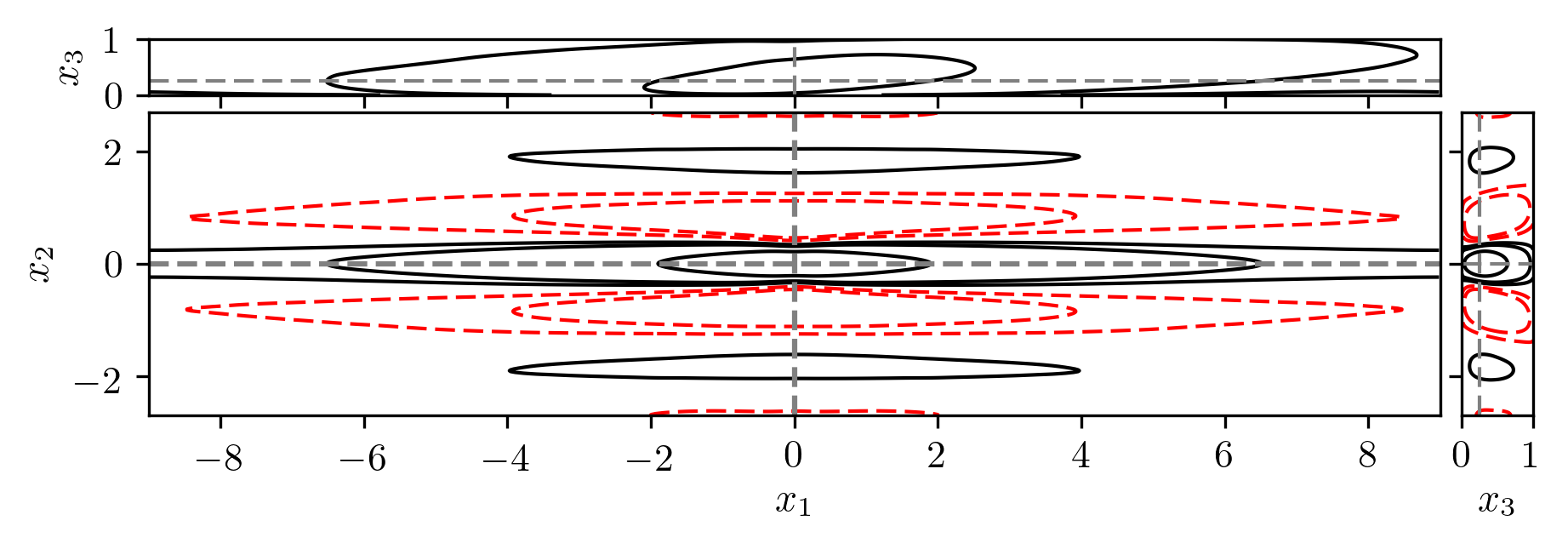}
			\caption{}
			\label{fig:Ng1} 
		\end{subfigure}
		
		\begin{subfigure}[b]{0.45\textwidth}
			\includegraphics[width=1\linewidth]{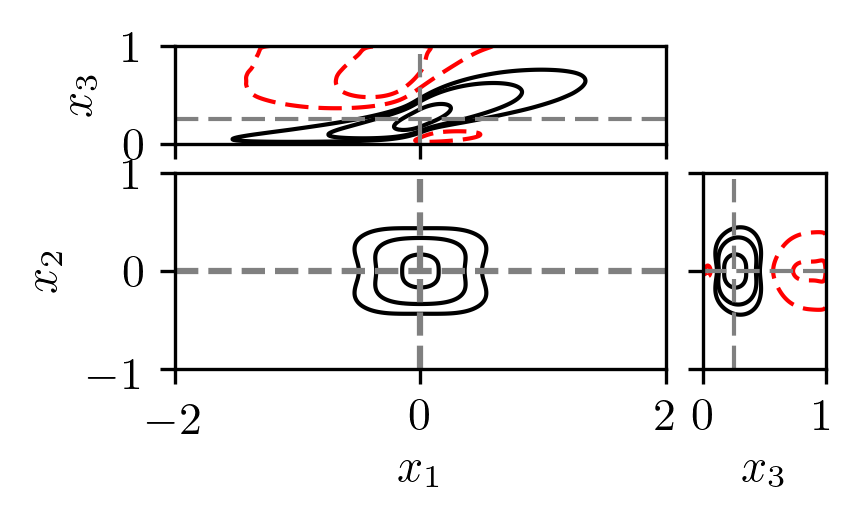}
			\caption{}
			\label{fig:Ng2}
		\end{subfigure}
		\begin{subfigure}[b]{0.45\textwidth}
			\includegraphics[width=1\linewidth]{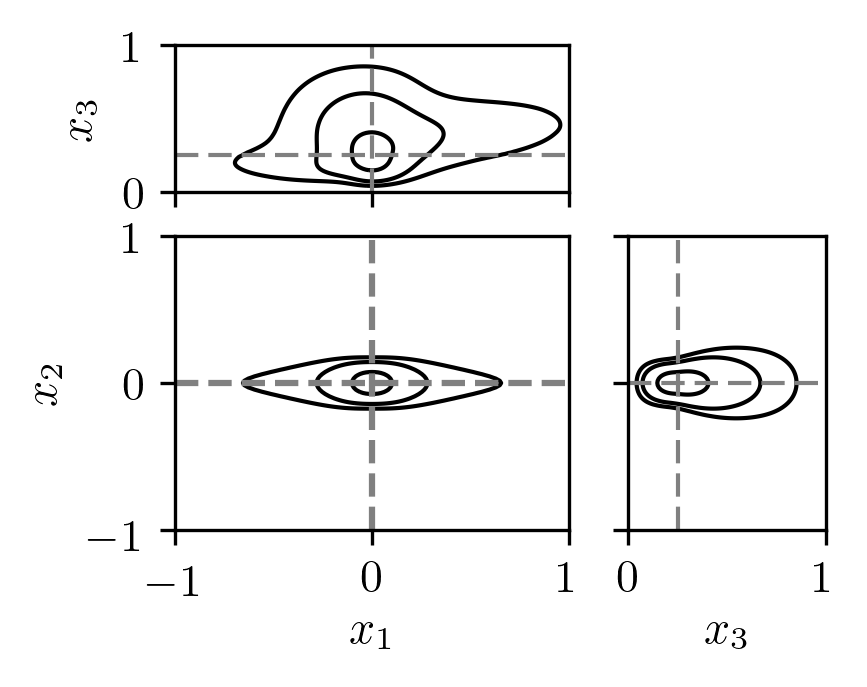}
			\caption{}
			\label{fig:Ng3}
		\end{subfigure}
		\caption[]{Visualisation of the two-point correlation $\mathsfi{C}_{ij}(\bs{x},\bs{x}')$ with reference point $\bs{x}'=[0,0,\frac{1}{4}L_3]$, each figure shows the contour lines of the horizontal $x_1$--$x_2$ cross-section at $x_3=\frac{1}{4}L_3$, the $x_1$--$x_3$ cross-section at $x_2=0$, and the $x_2$--$x_3$ cross-section at $x_1=0$. Figure (a), (b) and (c) respectively represent the $\mathsfi{C}_{11}$, $\mathsfi{C}_{22}$ and $\mathsfi{C}_{33}$ components. The contour lines are drawn for $\mathsfi{C}_{ij} = \lbrack -0.1, -0.05, 0.05, 0.1, 0.3\rbrack$.}
		\label{fig:corr_mat}
	\end{figure}
	
	\begin{figure} 
		\includegraphics[width=\textwidth]{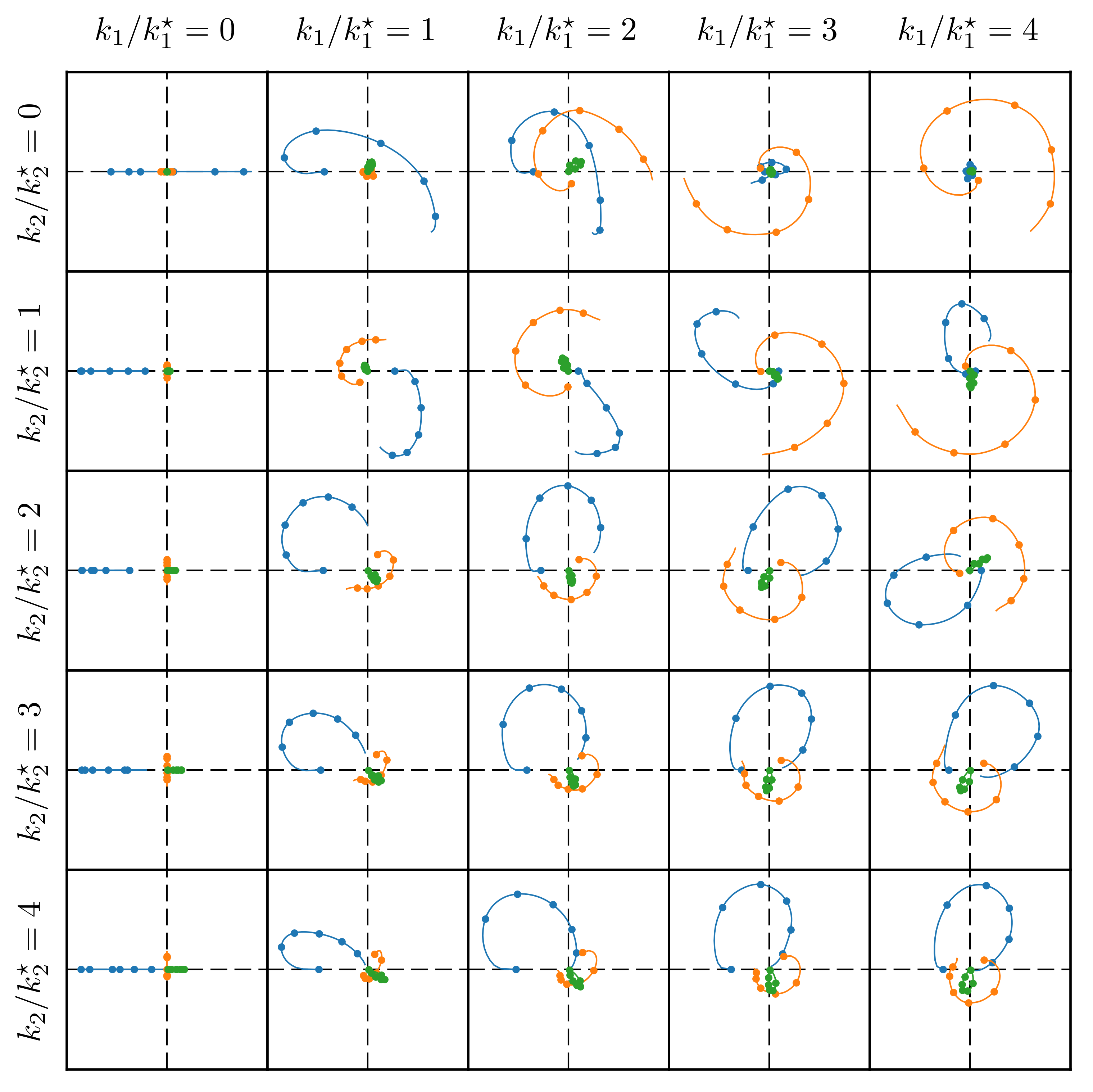}%
		\caption{Visualisation of $\widehat{\bs{\Psi}}_{\bs{k},1}(x_3)$, for $\left[k_1/k_1^\star, k_2/k_2^\star\right]= [0,0],\, \dots,[4,4]$, with $k_i^\star=2\upi/L_i$. Dots are placed at intervals $\Delta x_3=H/5$. The abscissas and ordinates are $\Real(\widehat{\bs{\Psi}}_{\bs{k},1}(x_3))$, $\Imag(\widehat{\bs{\Psi}}_{\bs{k},1}(x_3))$. Blue, orange, and green colours represent streamwise, spanwise and wall-normal directions, respectively.}
		\label{fig:eigenvector}
	\end{figure}

	We start by showing the structure of the covariance tensor. To this end, we visualise the correlation tensor, which is defined as
	\begin{align}
	\mathsfi{C}_{ij}(\bs{x},\bs{x}')=\frac{\mathsfi{B}_{ij}(\bs{x},\bs{x}')}{(\mathsfi{B}_{ij}(\bs{x},\bs{x})\mathsfi{B}_{ij}(\bs{x}',\bs{x}'))^{1/2}},
	\end{align} 
	and is an often-used nondimensionalised version of the covariance tensor \citep[see e.g. ][]{sillero2014two, jimenez2018coherent}). Figure \ref{fig:corr_mat} visualises different cross-sections of the diagonal components ($\mathsfi{C}_{11}$, $\mathsfi{C}_{22}$, $\mathsfi{C}_{33}$) of the correlation tensor for reference point $\bs{x}'=[0,0,\frac{1}{4}L_3]$. Figure \ref{fig:corr_mat} (a)  visualises $\mathsfi{C}_{11}$, which consists of  a central inclined positive lobe laterally surrounded by two negative correlated lobes. The streamwise velocity component of the tensor $\mathsfi{C}_{11}(\bs{x},\bs{x}')$, has been studied in \citet{fang2015large}, but almost exclusively for 1D streamwise and spanwise variations of $\bs{x}-\bs{x}'$. In this work, a similar decay of $C_{11}(\bs{x},\bs{x}')$ was found for the streamwise direction of $\mathsfi{C}_{11}(\bs{x},\bs{x} \pm 5H\bs{e}_1)=0.1$. Note that in \citet{fang2015large} at around $10H$ a zero crossing of the correlation was found and beyond this region anti-correlation, is found which stretches to at least to $30H$. However, our domain size is too small to observe this phenomenon. Moreover, an influence of periodic boundary conditions exists and leads to an overestimation of the correlation near the edges of the domain. The correlation of the spanwise velocity component $\mathsfi{C}_{22}$ has a main lobe, which has a steeper inclination angle compared to $\mathsfi{C}_{11}$, and has vertically situated anti-correlated side lobes. Finally, the vertical velocity component correlation $\mathsfi{C}_{33}$ has a relatively narrow main lobe in the spanwise direction. In contrast with the other components, no significant anti-correlated lobes are found. 
	
	Figure $\ref{fig:eigenvector}$ visualises the most energetic complex eigenfunctions $\widehat{\bs{\Psi}}_{\bs{k},1}(x_3)$ for the first 5 streamwise $k_1$ and spanwise $k_2$ wave numbers. Each eigenfunction can by an inverse Fourier transform be transformed to a 3D solenoidal vector field.

	\subsection{Turbulent flow field reconstruction}

	\begin{figure}
		\begin{subfigure}[a]{\textwidth}
			\includegraphics[width=1\linewidth]{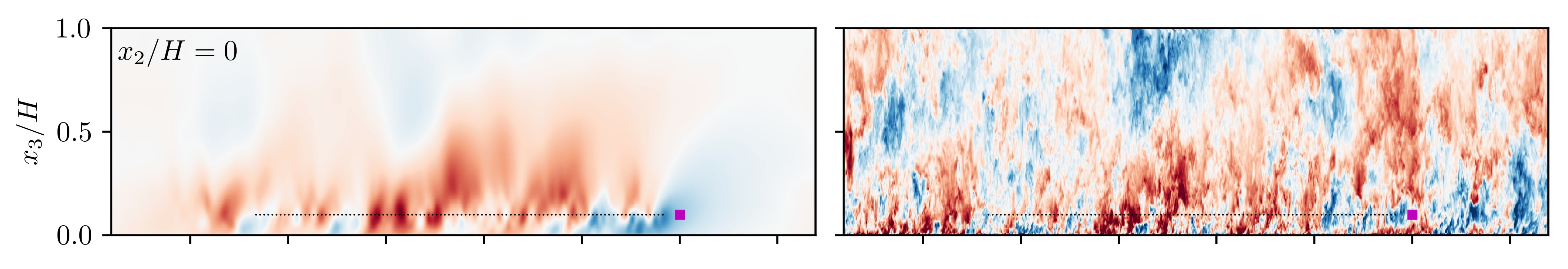}
			\includegraphics[width=1\linewidth]{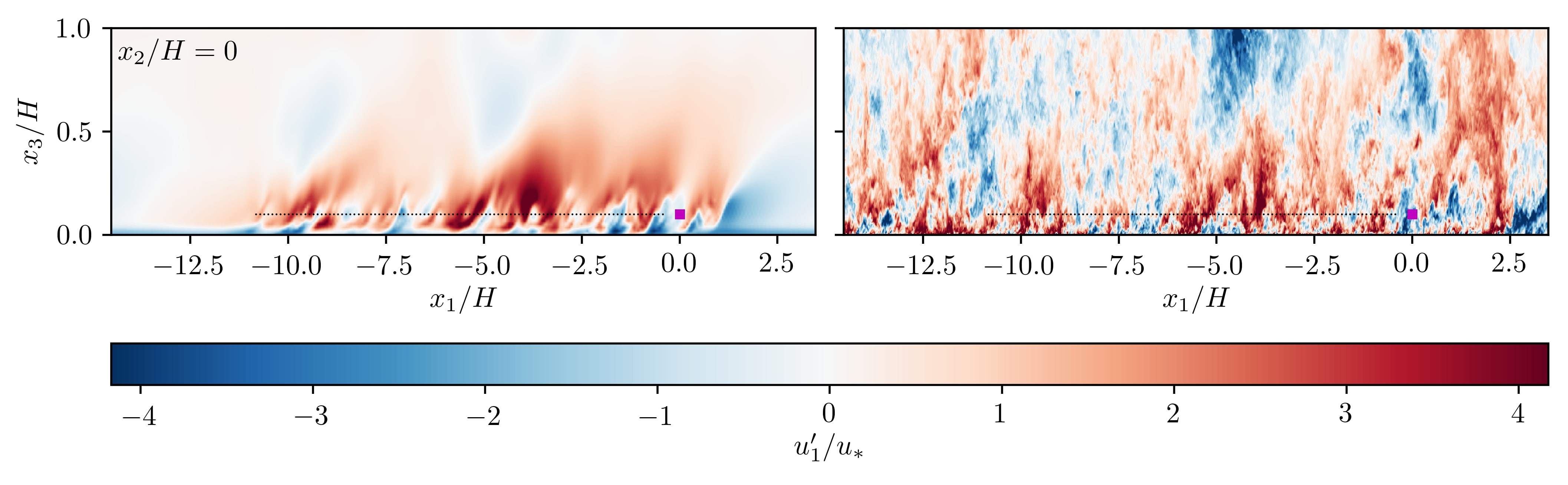}
			\caption{}
		\end{subfigure}
		\begin{subfigure}[]{\textwidth}
			\includegraphics[width=1\linewidth]{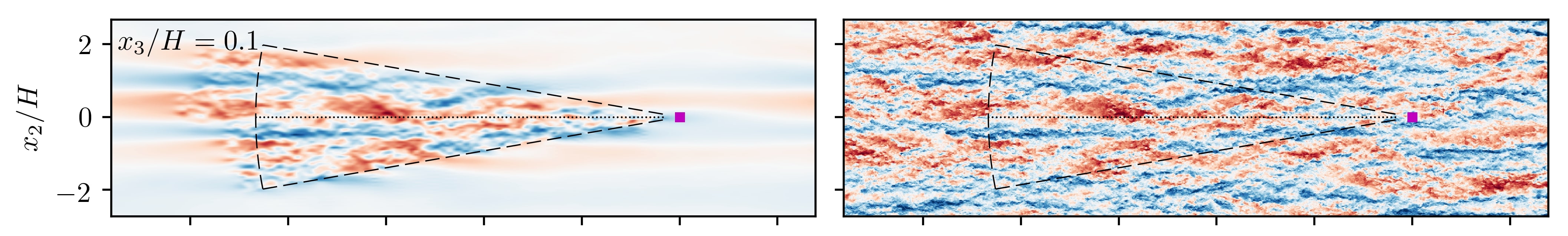}
			\includegraphics[width=1\linewidth]{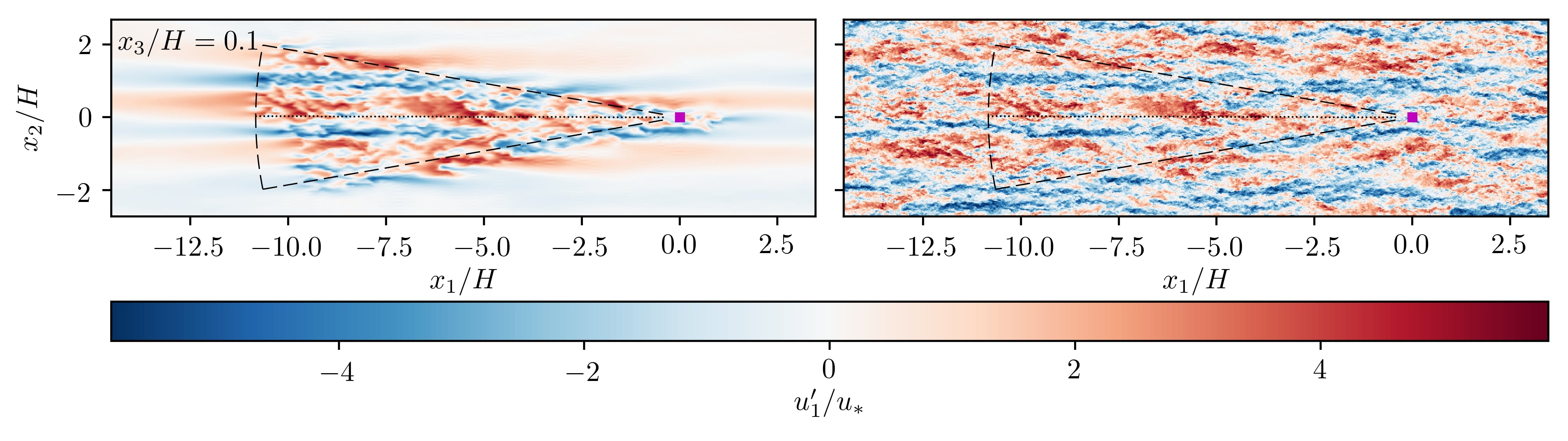}
			\caption{}
		\end{subfigure}
		\caption[]{The top (a) and bottom (b) figure respectively show the $x_2=0$ and $x_3=0.1H$ cross-section of the streamwise component of the velocity field fluctuations. The right and left hand side respectively show the reference field $u_1$ and the reconstructed field $\ft{u}_1$. The top and bottom subfigure respectively show the velocity field at $t=\ti$ and $t=\tf$. The purple square denotes the location of the lidar sensor. The lidar range gate centres are shown as dots. The dashed area denotes the total scanned area of the lidar.  }
		\label{fig:vel_sweepinglidar}
	\end{figure}
	
	\begin{figure}
		\begin{subfigure}[a]{\textwidth}
			\includegraphics[width=\textwidth]{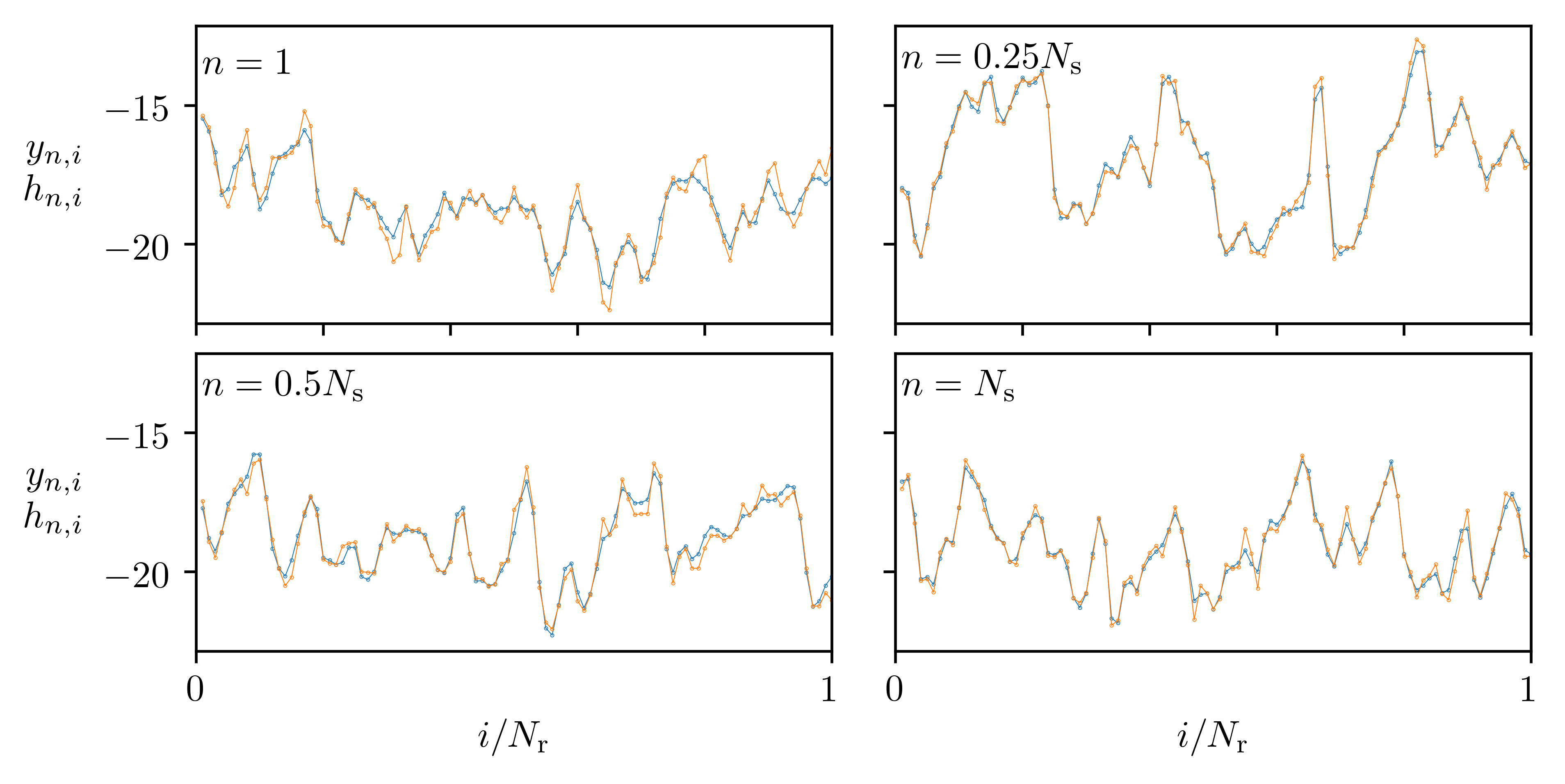}
			\caption{}
		\end{subfigure}
		\begin{subfigure}[a]{\textwidth}
			\includegraphics[width=\textwidth]{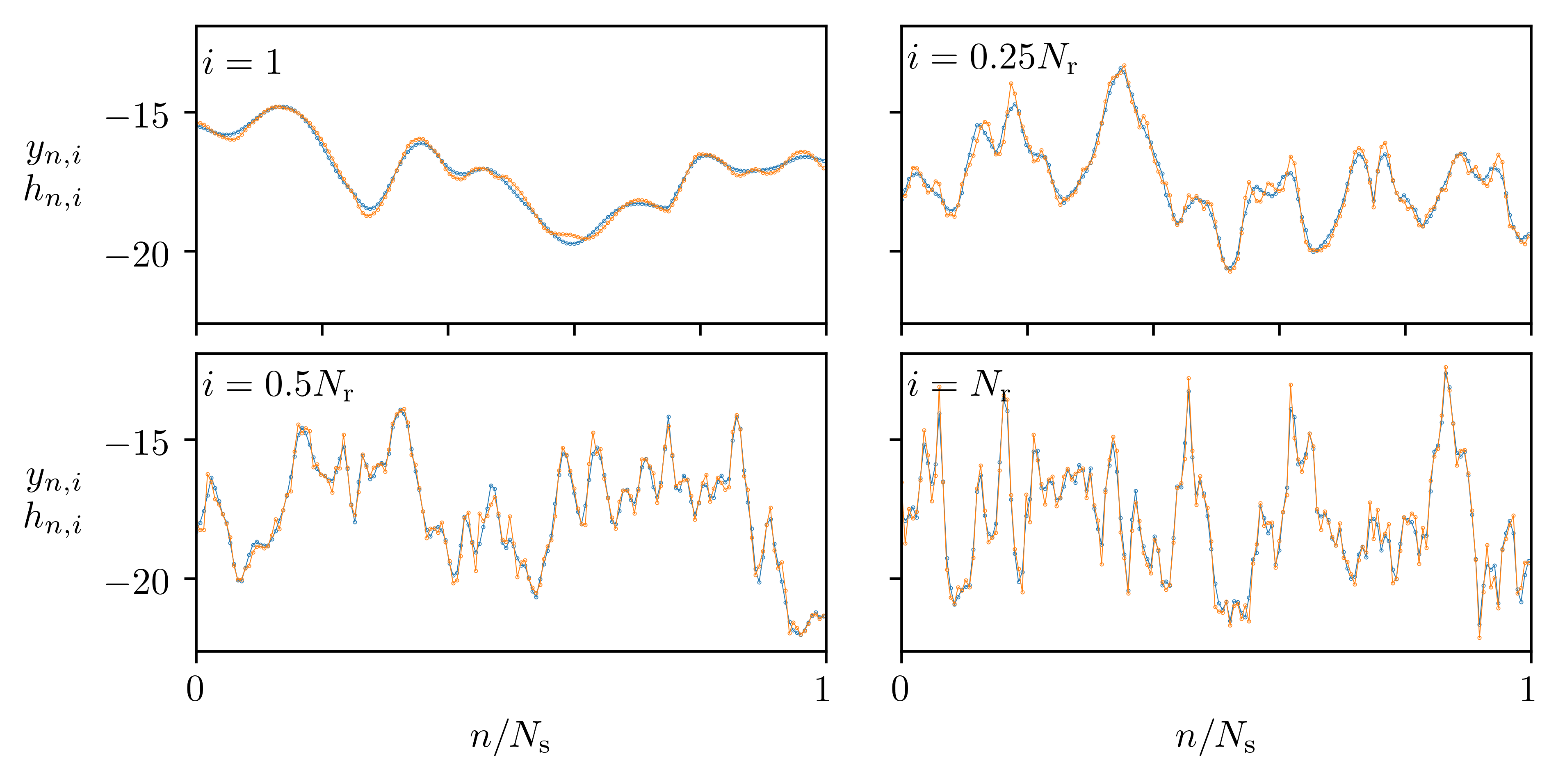}
			\caption{}
		\end{subfigure}
		\caption[]{Reconstructed ($h_{n,i}$, blue) and  reference ($y_{n,i}$, orange) measurements (a) for sample numbers $n=1$, $n= 0.25N_{\mt{s}}$, $n= 0.5N_{\mt{s}}$ and  $n= N_{\mt{s}}$ respectively and (b) for range gates $i=1$, $i= 0.25N_{\mt{r}}$, $i= 0.5N_{\mt{r}}$ and  $i= N_{\mt{r}}$. }
		\label{fig:lidar_vel}
	\end{figure}
	
	We now discuss the reconstruction of velocity fields from lidar observations. We start by looking at qualitative comparisons between the reference field and reconstructed velocity fields using an LES model. Subsequently, errors are quantified in more details, and comparison with reconstructions using a TFT model are also included.
	
	In figure \ref{fig:vel_sweepinglidar}, a comparison between the reconstructed velocity fluctuations $\ft{\bs{u}}'$ and the reference velocity fluctuations $\bs{u}'$ are visualised for the PPI lidar scanning mode. Cross-sections of the streamwise velocity are shown at the start ($t=\ti$) and end ($t=\tf$) of the observation time window. For the horizontal cross-sections (figure~\ref{fig:vel_sweepinglidar}b), in general a good correspondence between $\ft{\bs{u}}'$ and $\bs{u}'$ is found within the scanned area. We also observe that for the initial and final field, fluctuations in an area upstream and downstream of the scanning region, are also well reconstructed. This is explained by convective transport of flow information out of the measurement area during the measurements. Beyond this convective transport of flow information, additional upstream information is provided due to the relatively long $u_1$ correlations in streamwise direction. For the vertical cross-section (figure~\ref{fig:vel_sweepinglidar}a),  direct flow field observations are only available at $z=0.1H$. Due to the lack of mean transport in the vertical direction, additional information is only available due to the regularisation term, and the spatial coherence introduced by the LES model. Nevertheless, it is seen that the large scale motions are reasonably well constructed in the vertical plane.
	
	Figure~\ref{fig:lidar_vel} shows a comparison between reconstructed $\bs{h}_n$ and reference $\bs{y}_n$ measurements. We note that the optimisation problem effectively minimises this difference. Thus, as may be expected, the trends are well represented by the reconstructed measurements. The lidar measurement signal as a function of time  for a fixed range gate number (figure~\ref{fig:lidar_vel}b) transforms from a relatively smooth signal close to the sensor location to a more irregular further away due to the larger distance covered by the range gate.

\begin{figure}
		\includegraphics[width=\textwidth]{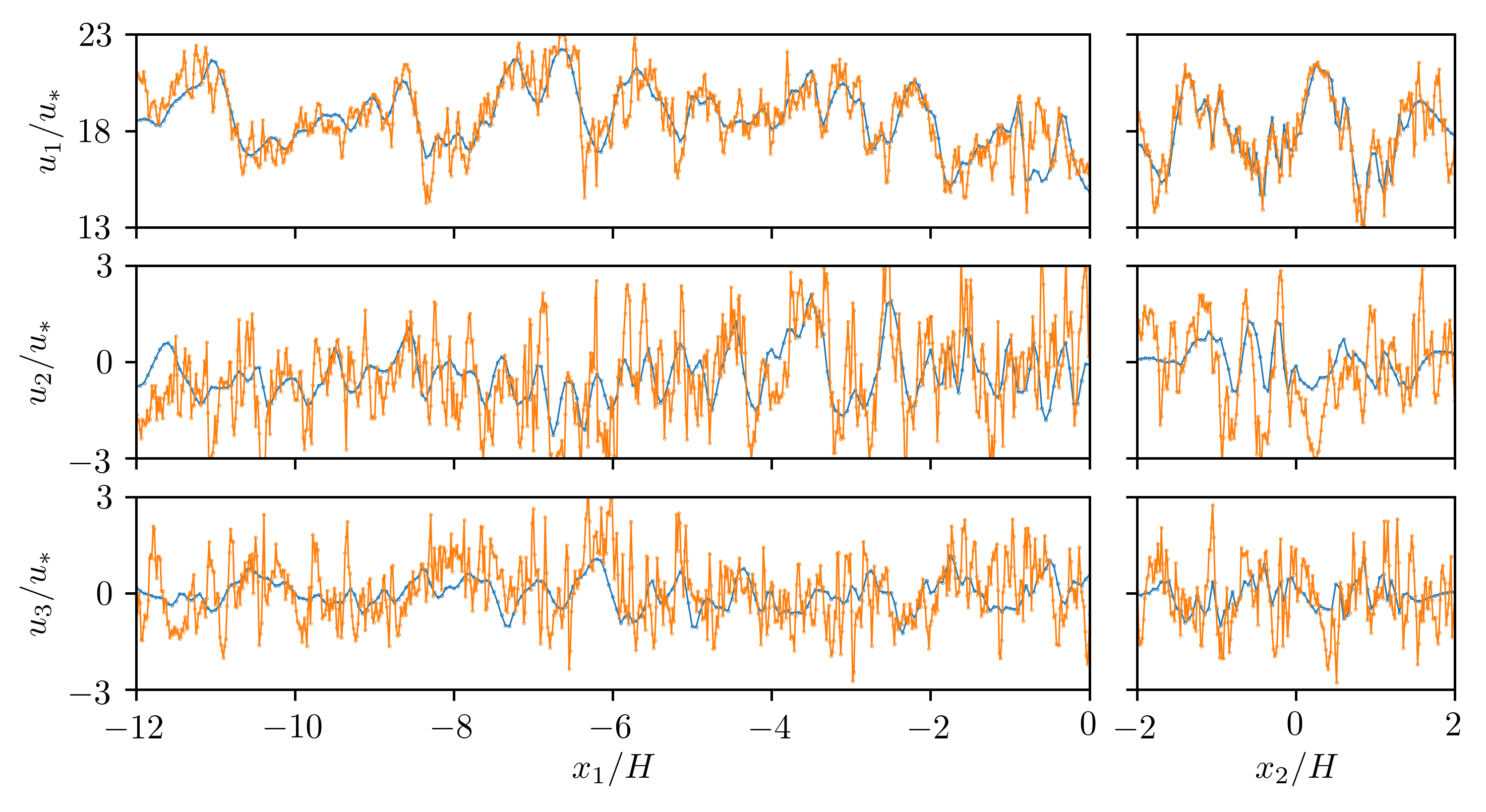}
		\caption[]{The top, middle and bottom figure present the streamwise, spanwise and wall-normal velocity component at $t = t_i + T/4$. The left and right hand side respectively show the velocity component along a line in the streamwise direction through the lidar mount, i.e. $(x_1,0,0.1H)$ and in the spanwise direction at mount height, and $8H$ upstream of the mount point, i.e. $(-8H,x_2,0.1H)$.  }
		\label{fig:cross_sections}
	\end{figure}

	Next, we focus on the reconstructed velocity field in the lidar plane, but at a moment in time when the lidar is pointing elsewhere. To this end, figure \ref{fig:cross_sections} compares the reconstructed velocity field to the reference field along two lines in the the streamwise and spanwise direction respectively, at time $t=\ti + T/4$. At this moment, the lidar is in a spanwise extremum position. It is observed that the streamwise velocity component $u_1$ is well constructed, except for the smallest scales. This is expected, since the lidar observations themselves  are spatially filtered, while the LES reconstruction mesh is also coarser than the reference. Looking at the streamwise and wall-normal components, the quality of the reconstruction appears somewhat lower, though the main trends are still recovered. However, when quantifying the errors in more detail (see below), we find that errors in absolute value are of the same order of magnitude for the three components, indicating that the reconstruction is of similar quality for all three directions, even though the lidar only measures along it's line of sight, which is dominantly oriented along the $u_1$ direction.
	
\begin{figure} 
		\includegraphics[width=\textwidth]{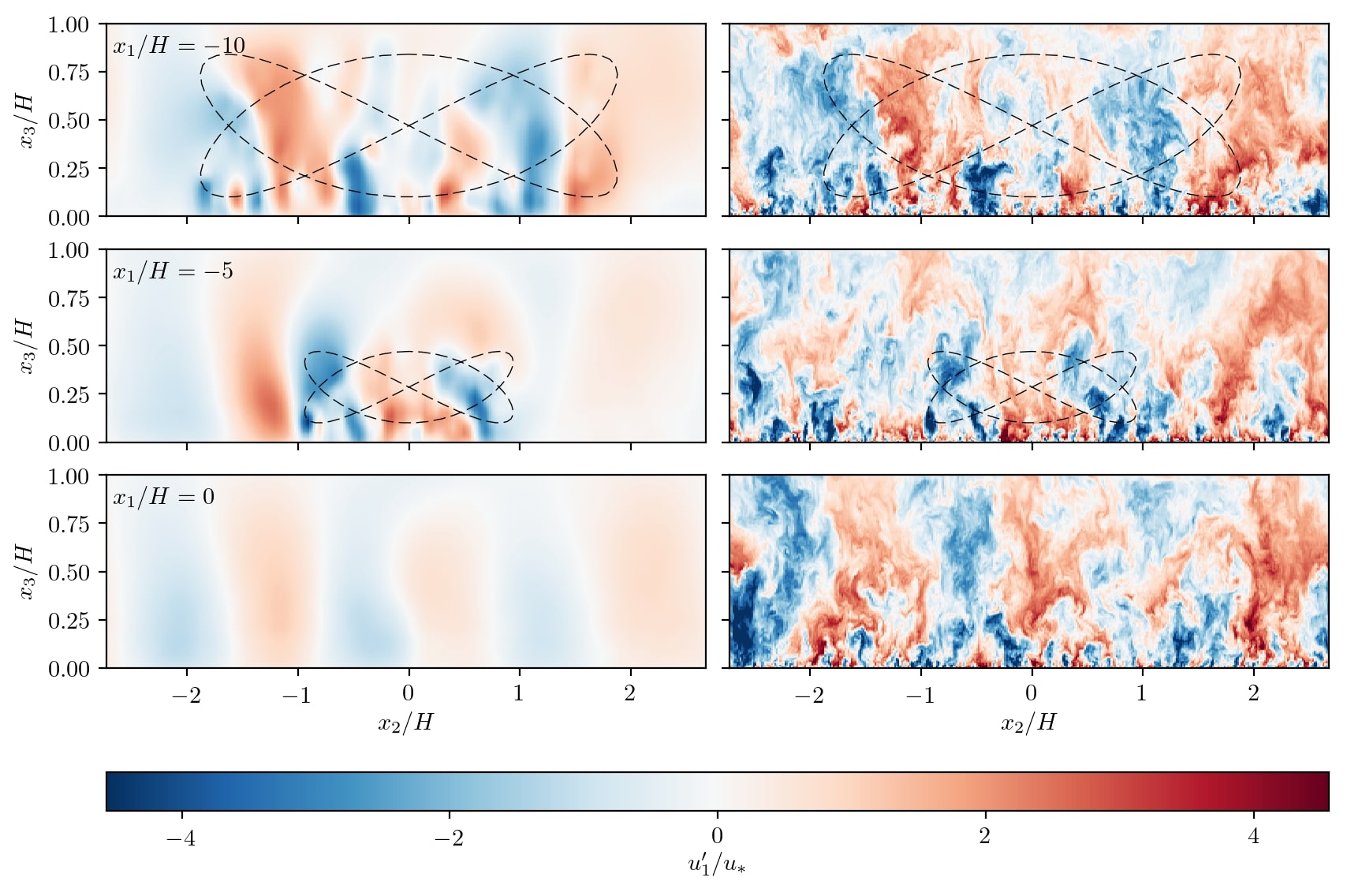}%
		\caption{Cross-section at $x_1 = -10H$, $x_1 = -5H$ and $x_1 = 0$ of the streamwise component of the velocity field at $t=\ti$. The left and right hand side respectively show the reconstructed velocity field $\ft{u}_1$ and reference velocity field $u_1$. The dashed line denotes the intersection of the lidar beam with the respective plane for the complete time horizon.}
		\label{fig:BL_field_Lissajous_yz_init}
	\end{figure}

	In Figure \ref{fig:BL_field_Lissajous_yz_init} the velocity field reconstruction for the Lissajous scanning mode is shown. Three spanwise--vertical planes are shown at $t=\ti$ comparing reconstructed to reference velocity field. The dashed line shows the trajectory of the intersect of these planes with the lidar beam during the complete time horizon. The further upstream the plane, the more spatially extended this trajectory is. As further quantified below,  quality of reconstruction along the height of the boundary layer is a lot better is for this scanning pattern, since direct measurements are available up to $0.9H$.
	
	\begin{figure} 
		\includegraphics[width=\textwidth]{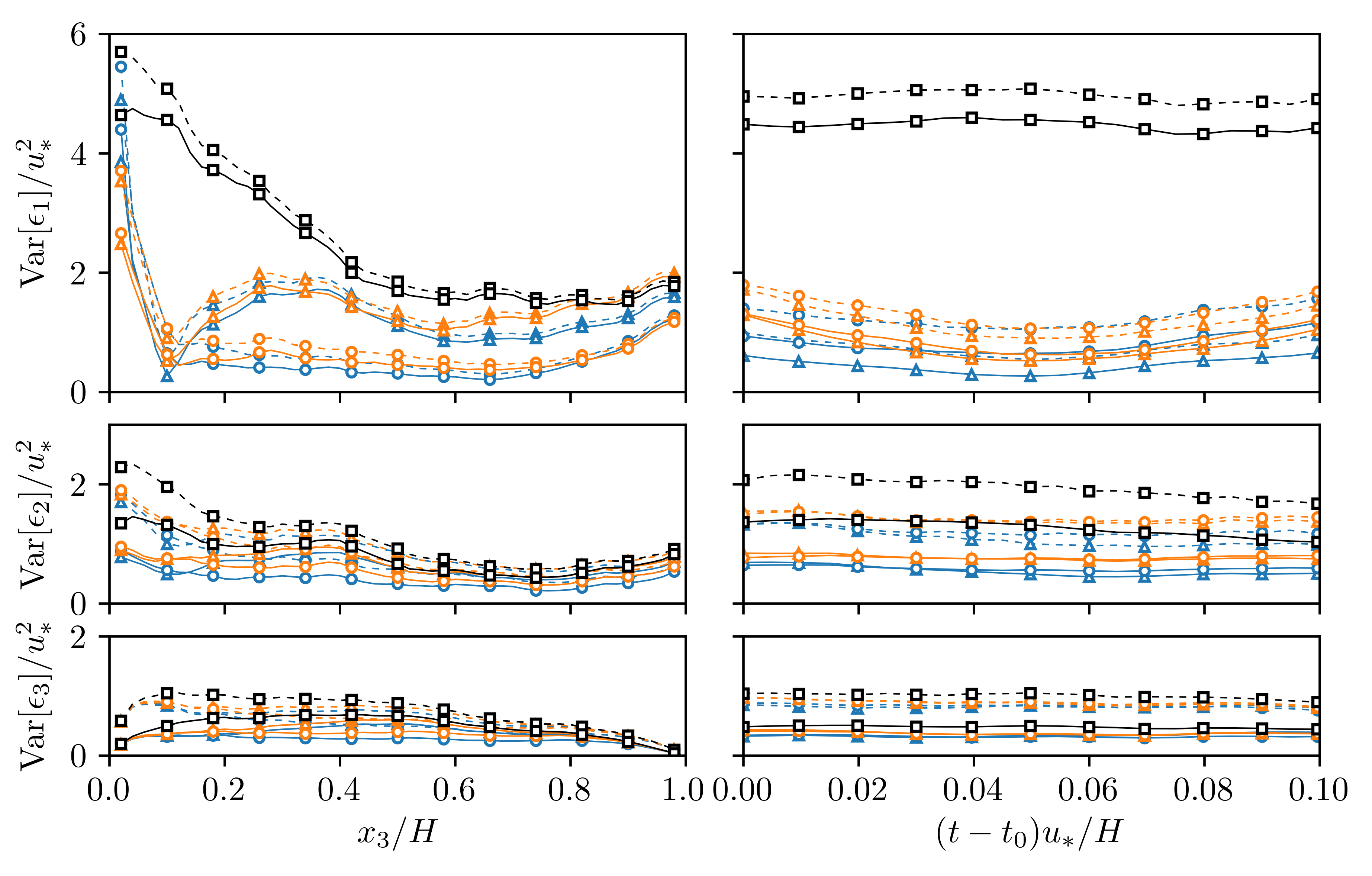}
		\caption{(left) Normalised error variances as function of height. (right) Evolution of error variances over the time window, evaluated at height $z=0.1H$. (top) Streamwise velocity component; (middle) spanwise velocity component; (bottom) vertical velocity component. ($\cdots$): error with respect to unfiltered reference; (---): error with respect to filtered reference. ($\triangle$): PPI scanning mode; ($\circ$): Lissajous scanning mode. (Blue): reconstruction based on LES, (Orange): reconstruction based on TFT, (Black, $\square$): background variance.}
		\label{fig:err_plane}
	\end{figure}

	In order to quantify errors in more detail, we define the error on the streamwise component as $\epsilon_1(\bs{x},t) = I_c^F \circ \widetilde{u}_1(\bs{x},t)-u_{1}(\bs{x},t)$, where $\widetilde{u}_1$ is the reconstructed velocity, $I_c^F$ is a coarse-to-fine interpolation operator (for which we simply use linear interpolation), and $u_{1}$ the fine-grid reference  velocity. We introduce the error variance averaged over a horizontal region of interest $\Gamma$ as $\mt{Var}[\epsilon_1(z,t)] = \langle (\epsilon_1)^2\rangle_{\Gamma}$. To this end, we select a horizontal region that is located inside the lidar observation area. To avoid boundary effects on the error, we omit regions that are less than one convective distance $U_{\infty}T$ away from the scanning boundaries, and use $\Gamma(z)=\mathsfbi{Q}_3(\upi)[r\cos(\phi) + U_\infty T, r\sin(\phi) ,z]$ with $r\in[R_0 ,R_{\mt{max}}-2U_\infty T]$, $\phi\in[-\Delta \phi/2, \Delta \phi/2]$ (see also \S\ref{ss:scanning_trajectories}). Similar to $\mt{Var}[\epsilon_1(z,t)]$, the error variances for the velocity components in spanwise and wall-normal direction are also introduced. Finally, since lidars effectively measure a filtered velocity field, we also construct errors based on a filtered reference velocity field. To this end, we filter the fine-grid LES field using a stream-wise cutoff filter with $k_{\mt{c}} = \upi/(\Delta p^2 + \Delta R^2)^{1/2}$, with $\Delta p^2= 3/(2\log 2) \Delta p_{1/2}^2$ roughly approximating the lidar filter kernel (\ref{eq:lidar_kernel}). 
	
	An overview of the normalised error variance is shown in figure~\ref{fig:err_plane} for reconstruction with an LES model, and as point of comparison also for reconstruction with the TFT model. Errors based on unfiltered and filtered reference fields are shown, and results for both the PPI and Lissajous scanning modes are included. As an additional reference, we also show, per component in figure~\ref{fig:err_plane}, the background variance $\langle u_{i}^{\prime2}\rangle_{\Gamma}$, which is the error obtained by predicting with the mean velocity profile. This is the best estimate without or far away from any measurements.
	
	
	First of all, it is observed in figure~\ref{fig:err_plane} that both scanning trajectories give good results at lidar mount height, though the PPI mode slightly outperforms the Lissajous mode, with a variance of the $u_{1}$ component that is on average only $15\%$ of the background variance, compared to $25\%$ for the Lissajous mode. The best results are obtained in the middle of the assimilation window, and progressively increases towards the bounds. When looking at higher altitudes, the Lissajous mode clearly outperforms the PPI mode, with an average normalized error variance of $25\%$ in the region $z$ from $0.1$ to $0.9$, compared to $55\%$ for the PPI mode. Moreover, in the lidar scanning region, the variance of errors in $u_{1}$ is of the same order than those of errors in $u_{2}$, and $u_{3}$ (in absolute value). This indicates that the estimation distributes the errors evenly in all direction. However, since in boundary layers, the background variance of $u'_{1}$ is much larger than that of $u'_{2}$, and $u'_{3}$, the relative error of the reconstruction in the $x_1$ direction is much lower. This explains, the better matching of the $u'_{1}$ signal in figure~\ref{fig:cross_sections} (which is simply larger in amplitude than the other components). Finally, it is further also found that LES consistently outperforms TFT, for example for the PPI scanning mode at lidar mount heigth, the error variance is $24\%$ an increase by $60\%$ compared to the LES. The maximum error of the TFT model in the scanning region is $35\%$, which is an increase by $70\%$ compared to the LES case.

	\section{Conclusion}\label{sec:conclusion}
	In the current study we investigated reconstructing turbulent flow field from lidar measurements by using a  4D-Var approach in combination with a LES model. The problem was regularised using the background covariance tensor, and reformulated using a POD basis. This allowed the elimination of the continuity constraint, and also led to a better conditioned formulation. Moreover, we used horizontal homogeneity of boundary layers to efficiently construct and represent the POD basis. In order to test the methodology, we constructed virtual lidar measurements from a fine-grid pressure-driven boundary layer, and reconstructed the turbulent flow field using LES on a coarser mesh. This allowed for a detailed error analysis. Different lidar scanning modes were investigated, and a comparison with a TFT model was also included. Overall, LES based reconstruction was quite effective. Inside the general lidar scanning region, we found that errors in the streamwise velocity fluctuation lie between $15\%$ and $25\%$ (error variance normalised by background variance). Moreover, LES outperformed TFT by $30\%$ to $70\%$. In spanwise and wall-normal directions, the reconstruction quality was the same in absolute value, but worse in relative value, since in turbulent boundary layers the streamwise background variability is substantially higher than spanwise and wall-normal variability. 
	
	In the current work, we studied a pressure driven boundary layer as a proxy for a neutral atmospheric boundary layer. In reality, effects related to the Ekman spiral (changing wind direction with height), and the capping inversion and free-atmosphere stratification can play an important role in boundary-layer height velocity profile. Moreover, near-wall stratification has strong effects on turbulence, requiring the use of correlation tensors that depend on the stratification regime. These are important directions for future research.
	
	A further working assumption in the current work, was the use of horizontal homogeneity for the construction of a POD basis in combination with an outer-scaling argument for scaling independent from surface roughness. In reality effects of inhomogeneity in terrain can have an effect on the outer region of the ABL, and further research needs to investigate whether an imperfect (based on homogeneity assumptions), but efficient POD basis remains effective when terrain features (e.g. wind turbines, buildings) are introduced. Moreover, the explicit inclusion of model uncertainties \citep{tr2006accounting}, may also further improve the methodology.
	
	\appendix
	
	\section{Derivation and validation of the adjoint gradient} \label{sec:adjoint_deriv}
	In this appendix, we derive the adjoint equations for the calculation of the gradient of the cost function $\J(\bs{a})$ to the control variables $\bs{a}$. For the derivation we use a Lagrangian approach \citep[see e.g.][]{borzi2011computational}, similar to the approach by \citet{goit2015optimal} to which we refer for further details. To this end, we first reformulate the optimisation problem (\ref{eq:optprob}) by removing the explicit solution operator $\mathcal{M}_{t}$, and instead explictly adding the state space constraints,  leading to
	\begin{align}
	&\underset{\bs{a}, \uvecfi, p}{\mt{minimise}}  &&\mathcal{J}(\bs{a}, \uvecfi) =  \frac{1}{2}\|\bs{a}\|^2 +  \frac{1}{2\gamma^2} \sum_{n=1}^{N_{\mt{m}}} \left\|\bs{y}_n-\bs{h}\left(\bs{u}(t), t_n\right) \right\|^2, \nonumber\\
	&\text{subject to} && \frac{ \partial \uvecfi}{\partial t}  +\uvecfi\cdot\bs{\nabla} \uvecfi  -u_*^2 H \bs{e}_1+\frac{1}{\rho} \bs{\nabla} \pfil - \bs{\nabla}\cdot\boldsymbol{\tau}_{\mt{SGS}} = \bs{0} \nonumber\\
	& && \bs{\nabla}\cdot\uvecfi = 0 \nonumber\\ 
	& && \uvecfi(\bs{x}, \ti)-\bs{\Psi}\bs{\Lambda}\bs{a} = \bs{0}.
	\label{eq:optprob_v2}
	\end{align}
	We note that by construction,$\J(\bs{a})=\mathcal{J}(\bs{a},\mathcal{M}_{t}(\bs{u}_0))$.
	For ease of notation the state variables and the adjoint variables are grouped together and respectively given by $\qvec=[\uvecfi, \pfil]^\top$ and $\qveca=[\uvecfia,\pfila, \bs{\chi}]^\top$. In an analogous way we group together the state space constraints $\bs{\mathcal{B}}(\bs{a},\bs{q})=[\NS^{\mt{m}},\NS^{\mt{c}},\con]^\top$, which are respectively the momentum, continuity equations and the constraint for the initial condition. The Lagrangian of above problem is now defined as $\L(\bs{a}, \qvec,\qvec^*) \triangleq \mathcal{J}(\bs{a},\qvec) + (\qveca,\bs{\mathcal{B}}(\bs{a},\qvec))$ and is given by
	\begin{align}
	\L(\bs{a}, \qvec, \qveca) &=\underbrace{\frac{1}{2}||\bs{a}||^2 + \frac{1}{2\gamma^2}\sum_{n=1}^{N_{\mt{s}}}||\bs{y}_n-\bs{h}_n || }_{\mathcal{J}} \nonumber \\
	& + \underbrace{\intt\int_{\Omega}\uvecfia\cdot\left(\frac{ \partial \uvecfi}{\partial t}  +\uvecfi\cdot\bs{\nabla} \uvecfi -u_*^2 H \bs{e}_1 +\frac{1}{\rho} \bs{\nabla} \pfil  - \bs{\nabla}\cdot\boldsymbol{\tau}_{\mt{SGS}}\right)\ \mt{d}\xvec\mt{d}t}_{(\uvecfia,\NS^{\mt{m}})}\nonumber\\
	&+ \underbrace{\intt\int_{\Omega}\pfila\left(\bs{\nabla}\cdot\uvecfi \right)\ \mt{d}\xvec\mt{d}t}_{(\pfila,\NS^{\mt{c}})}+\underbrace{\int_{\Omega} \bs{\chi}\cdot\left(\uvecfi(\bs{x},\ti)-\bs{\Psi}\bs{\Lambda}\bs{a}\right)\ \mt{d}\bs{x}}_{(\bs{\chi},\con)}.
	\end{align}
	It can be shown \citep[see e.g. ][]{troltzsch2010optimal} that, if the adjoint variables are chosen such that $\L_{\bs{q}}(\delta \bs{q}) = 0$ and the state space constraints are satisfied $\bs{\mathcal{B}}(\bs{a},\bs{q})=\bs{0}$, then $\J_{\bs{a}}(\delta\bs{a})=\L_{\bs{a}}(\delta\bs{a})$. Here we use the Riesz representation theorem to relate gradients to derivatives  \citep[see e.g. ][]{borzi2011computational}, for example for the cost function this gives
	\begin{equation}
	\J_{\bs{a}}(\delta \bs{a}) \triangleq\left. \frac{\mathrm{d}}{\mathrm{d}\alpha}\J(\bs{a}+ \alpha \delta \bs{a})\right|_{\alpha=0}= \left(\nabla_{\bs{a}} \J , \delta \bs{a}\right).
	\end{equation} 
	
	Further elaborating $\L_{\bs{q}}(\delta \bs{q}) = 0$ gives
	\begin{align}
	\L_{\qvec}(\delta \qvec) = &\left(\frac{\partial\mathcal{J}}{\partial \uvecfi},\delta \uvecfi\right)+\left(\uvecfia, \frac{\partial \NS^{\mt{\mt{m}}}}{\partial\uvecfi}\delta\uvecfi\right)+\left(\uvecfia,\frac{\partial\NS^{\mt{\mt{m}}}}{\partial p} \delta\pfil\right) \nonumber\\
	+&\left(\pfila,\frac{\partial \NS^{\mt{\mt{c}}}}{\partial \uvecfi}\delta\uvecfi\right)+ \left(\bs{\chi},\frac{\partial\con}{\partial \uvecfi}\delta\uvecfi\right) = 0,
	\end{align}
	where the terms that are trivially zero are left out. Partial integration of the terms $(\uvecfia,[\partial \NS^{\mt{m}}/\partial\uvecfi]\delta \uvecfi)$ and $(\pfila,[\partial \NS^{\mt{c}}/\partial {\uvecfi}]\delta \uvecfi)$, and $(\uvecfia,[\partial\NS^{\mt{\mt{m}}}/\partial p]\delta\pfil)$ respectively lead to the unforced adjoint momentum equation and adjoint continuity equation. This is a standard derivation: for the DNS-equations it can be found in \citet{bewley2001dns}, the additional LES terms are derived in \citet{goit2015optimal}. The remaining nonzero terms will be discussed in the subsequent sections. 
	
	\subsection{Adjoint contribution of the observations}
	Linearisation of the cost function to $\uvecfi$ gives  
	\begin{align}
	\left(\frac{\partial\mathcal{J}}{\partial \uvecfi},\delta \uvecfi\right)
	&=\int_{\ti}^{\tf}\int_{\Omega}\frac{1}{\gamma^2T_{\mt{s}}}\sum_{i=1}^{N_{\mt{r}}}\sum_{n=1}^{N_{\mt{s}}}(y_{n,i}-h_{n,i})  \nonumber\\
	& \qquad \qquad \times\mathcal{G}_{\mt{l}}\left(\mathsfbi{Q}(t)(\bs{x}-\bs{x}_{i}(t))\right)\bs{e}_{\mt{l}}(t) \, \mt{H}\!\left(\frac{T_{\mt{s}}}{2}-\left|t - t_{n-\frac{1}{2}}\right|\right)\cdot\delta \uvecfi\, \mt{d}\xvec\mt{d}t\nonumber\\
	&=\left(\sum_{i=1}^{N_{\mt{r}}}\bs{f}_{i},\delta \uvecfi\right),
	\end{align}
	which leads to the forcing term $\bs{f}_i$ for each lidar measurement in the adjoint momentum equations (\ref{eq:adjmom}).
	
	\subsection{Adjoint contribution of POD constraints} \label{a:PODconstraint}
	Linearisation of $(\bs{\chi},\con)$ to $\uvecfi$ gives 
	\begin{align}
	\left(\bs{\chi},\frac{\partial \con}{\partial \uvecfi}\delta \uvecfi\right) = \int_{\Omega} \bs{\chi}\cdot\delta \uvecfi(\bs{x},\ti)\ \mt{d}\bs{x}.\label{eq:PODconstraint}
	\end{align}
	The only other contribution at $\ti$ comes from the term $[\partial \delta\uvecfi/\partial t,\uvecfia]$ from the linearised momentum equation, and the combination of both needs to reduce to zero. This can be further elaborated by partial integration 
	\begin{align} \label{eq:consym}
	\int_{\ti}^{\tf}\int_{\Omega}  \uvecfia\cdot\frac{\partial \delta\uvecfi}{\partial t}\ \mt{d}{\xvec}\mt{d}t = \left.\int_{\Omega} \uvecfia\cdot \delta\uvecfi\ \mt{d}{\xvec}\right|_{\ti}^{\tf}-\int_{\ti}^{\tf}\int_{\Omega} \frac{\partial \uvecfia}{\partial t}\cdot \delta\uvecfi\ \mt{d}{\xvec}\mt{d}t.
	\end{align}
	The second term contributes to the adjoint momentum equations. The first term evaluated at $\tf$ can be eliminated if $\uvecfia(\bs{x},\tf)=\bs{0}$ is used as starting condition for the adjoint equations. The first term evaluated at $\ti$ combined with (\ref{eq:PODconstraint}) can be eliminated if $\bs{\chi}(\bs{x})=\uvecfia(\bs{x},\ti)$.
	
	\subsection{Derivation of the adjoint gradient}
	The linearisation of the reduced cost function $\J$ to the control variables $\bs{a}$, is given by (provided $\L_{\bs{q}}(\delta \bs{q}) = 0$ and $\bs{\mathcal{B}}(\bs{a},\bs{q})=\bs{0}$)  
	\begin{align}
	\J_{\bs{a}}(\delta \bs{a})=\L_{\bs{a}}(\delta \bs{a}) = \bs{a}^\top\delta \bs{a} - \int_{\Omega} \bs{\chi}\cdot\bs{\Psi}\bs{\Lambda}^{1/2}\delta\bs{a}\ \mathrm{d}\bs{x}.
	\end{align}
	The partial derivative to the $i$-th mode is readily identified as $\partial \J/\partial a_i = a_i - \lambda_i^{1/2}\int_{\Omega}\uvecfia(\bs{x},\ti)\cdot\bs{\psi}_i \ \mathrm{d}\bs{x}$ (for which $\bs{\chi}(\bs{x})=\uvecfia(\bs{x},\ti)$ was also used).  
	
	\section{Adjoint gradient validation} \label{sec:adjoint_valid}
	In this appendix we compare the adjoint gradient, with finite differences. The finite difference approximation of the Gâteaux derivative is given by
	\begin{equation}
	\left(\nabla \J,\delta \bs{a}\right ) \approx \frac{\J(\bs{a}+\alpha \delta \bs{a})-\J(\bs{a})}{\alpha}, \label{eq:fdevalgrad}
	\end{equation}
	where $\alpha$ is a factor which controls the step size. $\alpha=10^{-6}$ is found as a tradeoff between nonlinear effects $\sim\alpha ^2$ and the finite precision arithmetic by which the calculations are performed, starting to dominate at very small $\alpha$. The gradient is validated for both a laminar and turbulent initial velocity field $\ft{\bs{u}}$ of the reconstruction model. Since validating for all the components of the gradient $\nabla \J$ would be to time consuming, we use the steepest descent direction  $\delta \bs{a} = \nabla \J / \|\nabla \J\|$ obtained from the adjoint approach for the evaluation of the finite-difference method  (\ref{eq:fdevalgrad}), and compare results only for this direction. 
	
	The results are shown in table \ref{tab:FDvsADJ} for the PPI scanning mode, and are similar for the Lissajous scanning mode. The relative precision for the gradient is around $10^{-3}$ for the LES and $10^{-8}$ for the TFT model. The difference is explained by the differences in spatial discretisation errors, which are significantly larger for the nonlinear terms of the LES, compared to the TFT. Single components of the gradient have also been used, with similar results but are not further reported here. 
	
	\begin{table}
		\centering	
		\begin{tabular*}{\textwidth}{@{\extracolsep{\fill}}l l l c c c}
			Model &$\uvec$ \rule{0pt}{4.0ex}\rule[-3.0ex]{0pt}{0pt}& $\ft{\uvec}\ $& $(\nabla\J, \delta \bs{a})_{\mt{FD}}$ & $(\nabla\J, \delta \bs{a})_{\mt{AD}}$ & $\Big| \frac{(\nabla\J, \delta \bs{a})_{\mt{FD}}-(\nabla\J, \delta \bs{a})_{\mt{AD}}}{(\nabla\J, \delta \bs{a})_{\mt{FD}}}\Big|$\\
			\hline
			\Tstrut LES& L & T & $\SI{4.6018e4}{}$&  $\SI{4.6031e4}{}$& $\SI{8.6e-4}{}$\\
			LES&T & T & $\SI{5.6341e4}{}$ & $\SI{5.6390e4}{}$& $\SI{2.8e-4}{}$\\
			TFT& L & T  &$\SI{4.1162e4}{}$ & $\SI{4.1162e4}{}$& $\SI{1.6e-8}{}$\\
			TFT&T & T &$\SI{5.0161e4}{}$ &$\SI{5.0161e4}{}$ & $\SI{2.7e-8}{}$\\
		\end{tabular*}
		\caption{Comparison of the adjoint and finite difference gradient, for LES and TFT models, using different initial states. `L' and `T' are respectively abbreviations for laminar and turbulent initial flow conditions.}
		\label{tab:FDvsADJ}
	\end{table}

\bibliographystyle{apalike}

\providecommand{\noopsort}[1]{}\providecommand{\singleletter}[1]{#1}%

\end{document}